\documentclass[aps,onecolumn,prd,preprintnumbers,superscriptaddress,nofootinbib,bibnotes,floatfix,
longbibliography,
]{revtex4-2}

\pdfoutput=1
\usepackage{amsmath}
\usepackage{amsfonts}
\usepackage{amssymb}
\usepackage{mathrsfs}
\usepackage{graphicx}
\usepackage{color}
\usepackage{bbding}
\usepackage{tabularx}
\usepackage[usenames,dvipsnames,table]{xcolor}
\usepackage[normalem]{ulem}
\usepackage{makecell}
\usepackage{siunitx}
\usepackage{enumerate}
\usepackage{multirow}
\usepackage{makecell}
\usepackage{upgreek}
\usepackage{comment}
\usepackage{subfigure}
\usepackage{booktabs}

\usepackage{marvosym}
\definecolor{darkred}{rgb}{0.5,0,0}
\definecolor{darkblue}{rgb}{0,0,0.5}
\definecolor{firebrick}{rgb}{0.75,0.125,0.125}
\definecolor{darkgreen}{rgb}{0,0.5,0}
\usepackage[colorlinks=true,linkcolor=firebrick,citecolor=darkgreen,urlcolor=darkblue]{hyperref}

\newcommand{\beq}{\begin{equation}}
\newcommand{\eeq}{\end{equation}}

\newcommand{\eq}{Eq.}

\graphicspath{{figures/}}

\newcommand{\orcid}[1]{\href{https://orcid.org/#1}{\includegraphics[width=10pt]{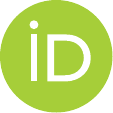}}}

\begin{document}

\title{Ultra High Energy Neutrino Event KM3-230213A as a Signal of Electroweak Vacuum
Turbulence in Merging Black Hole Binaries}

\author{Alexander S.~Sakharov \orcid{0000-0001-6622-2923}}
\email{alexandre.sakharov@cern.ch}
\affiliation{Department of Mathematics and Physics, Manhattan University,\\
4513 Manhattan College Parkway, Riverdale, NY 10471, United States of America}
\affiliation{Experimental Physics Department, CERN, CH-1211 Gen\`eve 23, Switzerland}

\author{Rostislav Konoplich \orcid{0000-0002-6223-7017}}
\email{rostislav.konoplich@manhattan.edu}
\affiliation{Department of Mathematics and Physics, Manhattan University,\\
4513 Manhattan College Parkway, Riverdale, NY 10471, United States of America}
\affiliation{Department of Physics, New York University,\\
726 Broadway, New York, NY 10003, United States of America}

\author{Merab Gogberashvili \orcid{0000-0002-7399-0813}}
\email{gogber@gmail.com}
\affiliation{Department of Exact and Natural Sciences, Javakhishvili Tbilisi State University,
Tbilisi 0179, Georgia}
\affiliation{ Department of High Energy Physics, Andronikashvili Institute of Physics,
Tbilisi 0177, Georgia}

\date{\today}
%\date{today}

\begin{abstract}
The recent detection of the ultra-high-energy neutrino event KM3-230213A ($\sim$220 PeV) by KM3NeT telescope poses a challenge to conventional astrophysical models, particularly in light of the absence of similar $\gtrsim$100 PeV events in IceCube data, despite its larger exposure. We propose a novel mechanism in which binary black hole mergers act as transient neutrino sources via gravitationally induced electroweak vacuum instability. In this scenario, the extreme spacetime curvature near the horizons during the final inspiral phase destabilizes the Higgs vacuum, triggering nucleation of true-vacuum bubbles. Collisions between these bubbles produce microscopic black holes that rapidly evaporate via Hawking radiation, emitting intense, short-lived bursts of neutrinos with energies exceeding 100 PeV. The resulting neutrino fluence follows a heavy-tailed distribution, allowing rare but highly luminous sources to account for events like KM3-230213A while remaining consistent with IceCube's non-detections. This framework links gravitational wave sources to ultra-high-energy neutrino production and suggests that future multi-messenger observations may detect electromagnetic signatures from microscopic black hole evaporation.
\end{abstract}

\maketitle

%%%%%%%%%%%%%%%%%%%%%%%%%%%%%%%%%%%%%%%%%%%%%%%%%%%%%%%%%%%%%%%%%%%%%%%%%%%%
%%%%%%%%%%%%%%%%%%%%%%%%%%%%%%%%%%%%%%%%%%%%%%%%%%%%%%%%%%%%%%%%%%%%%%%%%%%%

\section{Introduction}
\label{sec:intro}

The recent detection of the ultra-high-energy (UHE) neutrino event KM3-230213A~\cite{KM3NeT_220PeV} by the ARCA detector of the KM3NeT telescope has sparked significant interest within the astrophysics community. This event, identified through a track-like signature consistent with a muon of approximately 120~PeV, implies a parent neutrino with an energy of \(\sim 220\)~PeV and a 68\% confidence interval spanning 110–790~PeV, making it the most energetic neutrino ever observed.

The observation introduces a notable tension when compared with data from the IceCube Neutrino Observatory. Despite IceCube's significantly larger effective area and longer operational duration, it has not reported any neutrino detections above 10~PeV~\cite{IceCubePeV2}. Quantitative analyses reveal a $3.5\sigma$ discrepancy between the detection of KM3-230213A and the absence of comparable events in IceCube, assuming an origin in the diffuse isotropic neutrino background. Even under cosmogenic neutrino scenarios based on standard model expectations, the tension remains in the range of $3.1\sigma$ to $3.6\sigma$. Alternative hypotheses involving steady or transient point sources are somewhat less disfavored, but still exhibit tensions of $2.9\sigma$ and $2.0\sigma$, respectively~\cite{titans}.

To account for a cosmogenic origin of the KM3-230213A event~\cite{cosmogenicCL1,cosmogenicCL2,cosmogenicCL3}, the KM3NeT Collaboration argues that its detection can be accommodated within standard models only by extending the redshift integration of the cosmogenic neutrino flux up to $z \approx 6$, and by assuming that protons form only a subdominant component of the ultra-high-energy cosmic-ray (UHECR) population~\cite{KM3NeT:Cosmogenic1}. This scenario imposes stringent constraints on viable cosmic accelerators and must remain consistent with observational bounds from the Pierre Auger Observatory and the Telescope Array.
The possible origins of such cosmogenic neutrinos at the energy scale of KM3-230213A have been analyzed across several frameworks, including transient point sources, diffuse astrophysical emission, and line-of-sight interactions of UHECRs with cosmic background photons~\cite{CR1}. In the case of a cosmogenic point source, active galactic nuclei (AGNs) emerge as plausible candidates capable of producing such high-energy neutrino events~\cite{CR1}. For the transient UHECR source scenario, reproducing the KM3-230213A event would necessitate a new class of transients that are not only highly energetic and gamma-ray dark, but also significantly more abundant than those currently known~\cite{CR1}.
An alternative explanation involves cosmogenic neutrino emission from a nearby transient source, such as a gamma-ray burst (GRB), where the angular and temporal features of the resulting flux can dominate over the diffuse background under favorable intergalactic magnetic field conditions~\cite{cosmogenic1}.
A Galactic origin has also been considered~\cite{KM3NeT:Galactic1}, but the extreme energy of KM3-230213A poses a serious challenge to this scenario. It would require acceleration mechanisms within the Milky Way that greatly exceed the capabilities of known Galactic sources. Furthermore, the absence of coincident detections in other observatories further disfavors a Galactic interpretation.
A distinct hypothesis is presented in~\cite{Neronov:YearLongSource1}, where the event is attributed to a flare from an isolated transient source within the KM3NeT field of view, lasting less than two years. This model is compatible with existing observations from KM3NeT, IceCube, and the Pierre Auger Observatory. In such a scenario, the parent protons must exhibit a very hard energy spectrum above $10^{19}~\mathrm{eV}$, or the neutrino production must proceed through photohadronic interactions with ambient infrared photons.
In the broader context of the global neutrino sky, the KM3-230213A event remains an exceptional outlier~\cite{KM3NeT:Landscape1}. Its detection calls for a reassessment of current models and may signal the presence of previously unknown astrophysical sources or novel mechanisms responsible for UHE neutrino production.

An important constraint on the properties of potential sources of the KM3-230213A neutrino event arises from the associated gamma-ray emission, as examined in~\cite{KM3NeT:conterpart1}. Assuming that the observed neutrino was produced either within an astrophysical accelerator or through line-of-sight interactions of UHECRs with background photons, the expected gamma-ray flux resulting from ensuing electromagnetic cascades is computed. These cascades are initiated by co-produced UHE photons interacting with the extragalactic background light. The study explores a range of source redshifts and intergalactic magnetic field (IGMF) strengths and concludes that unless the IGMF is stronger than \( B \gtrsim 3 \times 10^{-13} \,\mathrm{G} \), or significant gamma-ray attenuation occurs within the source, the cascade-induced GeV--TeV gamma rays should be observable with current instruments. In the case of substantial internal gamma-ray suppression, the source is expected to be radio-loud. These considerations place meaningful constraints on both the nature and environment of the potential source and underscore the value of multi-messenger approaches in probing the origin of such UHE neutrino events. Extensive efforts have been made for looking for the association of blazars located within this 99\% confidence region of the arrival direction
of the KM3NeT's event~\cite{KM3NeT:conterpart1}. These candidates were identified based on their multiwavelength properties, utilizing archival data and dedicated observations. The study~\cite{KM3NeT:conterpart1} concluded that, while several blazar candidates exhibit properties consistent with potential neutrino emission, none could be definitively identified as the source of the UHE neutrino~\footnote{Further attempts to interpret blazars located in directional proximity to the KM3-230213A event as plausible UHE neutrino sources have been made in~\cite{blazar1,blazar2}, but these too remain inconclusive.}.
A systematic search is conducted in Fermi-LAT data for gamma-ray cascade signals coinciding spatially and temporally with the KM3-230213A event~\cite{cascade2}.  The analysis~\cite{cascade2} finds no statistically significant gamma-ray excess in any temporal or spatial bin that could be associated with the UHE neutrino event.

Therefore, from the current multi-messenger observational constraints, a favorable source of UHE neutrinos such as KM3-230213A must satisfy several stringent and mutually consistent requirements. First, the source must be capable of producing neutrinos with energies exceeding 100~PeV, indicating the presence of extremely efficient acceleration mechanisms that transcend conventional astrophysical processes, or potentially invoke physics beyond the Standard Model (SM). Furthermore, the source must be sufficiently rare and transient to avoid overproducing the diffuse neutrino flux already constrained by IceCube and the Pierre Auger Observatory, while still being detectable by KM3NeT/ARCA either due to a favorable location in the Southern Hemisphere or/and as a result of an exceptionally luminous burst. The observed absence of clustering or repetition implies that the event may originate from a unique, possibly stochastic process, such as a rare transient flare or a singular UHE neutrino burst.
Second, the source must permit a significant suppression of accompanying high-energy gamma-ray emission to remain consistent with the lack of correlated electromagnetic signatures reported by Fermi-LAT, H.E.S.S., MAGIC, LHAASO, HAWC, and other facilities including future CTA. Such suppression could arise from internal attenuation mechanisms within the source, interactions with dense photon fields, or due to extragalactic propagation effects like pair production on the extragalactic background light (EBL), followed by deflection and dispersion in intergalactic magnetic fields.
Finally, the source should evade existing limits from cosmic-ray observatories on associated UHECR fluxes, and be potentially observable in non-electromagnetic messengers or exhibit signatures in future multi-messenger campaigns, without violating bounds from cosmogenic neutrino models. This balance of high-energy neutrino production, electromagnetic invisibility, spatial and temporal uniqueness, and consistency with existing observables defines the framework within which viable theoretical models must operate.

A number of source models invoking physics beyond the Standard Model have been proposed to account for UHE
neutrino events such as KM3-230213A.
These include Hawking evaporation of primordial black holes (PBHs)~\cite{pbh1}~\footnote{Further study performed
in~\cite{pbh1NotReal} invalidates the conjecture of~\cite{pbh1}.},
as well as PBHs experiencing memory burden effects that enhance late-time particle emission~\cite{BHburden1,BHburden2}.
Scenarios involving five-dimensional PBHs within theories of large extra dimensions have also been considered~\cite{pbhExtraD1,pbhExtraD2}. In these models, PBHs predominantly emit gravitationally coupled modes, rendering most of their evaporation effectively invisible to observers confined to the SM brane. However, during the final stages of evaporation, a burst of SM particles may be emitted at temperatures approaching the characteristic quantum gravity scale of \(\sim 10^8\)~GeV, potentially producing detectable neutrinos. Another possibility involves PBHs carrying a small charge under a hidden U(1) symmetry, which can remain in a quasi-extremal state for extended duration. These black holes exhibit suppressed emission of lower-energy neutrinos while still allowing for the production of PeV–EeV neutrinos, consistent with observations~\cite{pbhExtrem1}. Decaying heavy dark matter (DM) particles constitute another class of potential sources~\cite{dm1,dm2,dm3,dm6}, with some scenarios linking DM production to PBH evaporation in the early Universe~\cite{dm4,pbhDM2}. Additional models consider sources emitting sterile neutrinos that convert to active neutrinos via oscillations amplified by new physics effects~\cite{nuSterile1}, either produced before recombination or generated through DM decay in the present epoch~\cite{nuSterile2}.
A distinct mechanism has been proposed to enhance the cosmogenic UHE neutrino flux, wherein free neutrons resulting from photodisintegration of heavy cosmic ray nuclei oscillate into mirror neutrons ($n \rightarrow n'$), which subsequently decay via mirror baryon-number-violating processes, transferring a significant portion of their energy into SM neutrinos~\cite{mirrorNeutron1}~\footnote{The idea of applying neutron–mirror neutron oscillations to UHECR propagation was first discussed in~\cite{mirrorNeutron2}.}.
Additionally, building on earlier developments~\cite{CnuB1,CnuB2}, coherent elastic neutrino nucleus scattering (CE$\nu$NS) of UHE cosmic rays on the cosmic neutrino background has been proposed as a viable mechanism for boosting relic neutrinos to PeV energies, potentially accounting for the KM3-230213A event~\cite{CEnuNS1}~\footnote{Scattering of UHE neutrinos with DM may also occur~\cite{NuDMscat1}.}.
Furthermore, alternative scenarios to standard neutrino signals propose that energetic DM particles interact within the Earth, upscattering into intermediate dark sector states that subsequently decay into muons, thereby offering a distinct channel for generating observable UHE muon tracks in neutrino telescopes~\cite{DMmuEarth1,DMmuEarth2}.

In this work, we propose a model in which the source of UHE neutrinos is a binary black hole (BBH) merger during which the Higgs vacuum enters a dynamically unstable and turbulent phase. This transient state facilitates rapid nucleation and collisions of vacuum bubbles over a brief interval near the merger's final stages~\cite{tribo}. The scenario builds on the hypothesis that the Higgs potential becomes destabilized in the strong gravitational field near black hole (BH) horizons, resulting in an enhanced decay rate of the metastable electroweak (EW) vacuum into a lower-energy true vacuum state~\cite{higgsStrumia1,higgsStrumia2}.
The metastability of the EW vacuum is supported by precise measurements of the Higgs boson and top quark masses, currently
reported as \( M_h = 125.20 \pm 0.11 \,\mathrm{GeV} \) and \( M_t = 172.56 \pm 0.31 \,\mathrm{GeV} \), respectively~\cite{PDG}. These values place the SM vacuum in a metastable region~\cite{smInstable1}, implying the possibility of a catastrophic transition to a lower-energy state in the future. As argued in Ref.~\cite{tribo}, the presence of an astrophysical BH can locally modify the effective Higgs potential, reducing the height, width and position of the potential barrier that stabilizes the EW vacuum. This local destabilization leads to an elevated nucleation rate of the true vacuum bubbles within a thin shell-like region situated just outside the BH horizon.
The resulting configuration is characterized by a turbulent medium composed of rapidly nucleating bubbles of the true vacuum interspersed within the metastable EW vacuum. Although bubble formation is continuously active in this region,
most bubbles are promptly swallowed by the BH horizon, rendering the metastable EW state
observationally intact to a distant observer in the single BH case.
However, in the final stages of a BBH merger, the gravitational influence of the two closely orbiting horizons can create a transient cavity between them. Within this gap, vacuum bubbles may become temporarily trapped, allowing them to grow, collide, and possibly percolate. These high-energy collisions can result in the formation of microscopic BHs (\(\mu\)BHs), via mechanisms described in~\cite{pbhBubble4, pbhBubble3}.
The \(\mu\)BHs generated in this way are expected to have very low masses~\cite{tribo}.
In particular, $\mu$BHs with masses below approximately half a ton are capable of
emitting neutrinos with energies exceeding 50~PeV, with their entire mass radiated away within a
timescale of nanoseconds.
The lighter the $\mu$BH, the harder the emitted neutrino spectrum and the shorter the burst duration.
More broadly, within this framework, the duration of Hawking radiation emission is constrained by the
characteristic timescale of the final inspiral and coalescence phase of a BBH merger, which, according to
LIGO, Virgo, and KAGRA observations~\cite{bbhLIGO1, bbhLIGO2, bbhLIGO3, LVKrun3},
is typically shorter than one second.
Once the merger completes and a single horizon forms, the transient conditions that permitted the
EW vacuum destabilization cease to exist.
All remnants of the true vacuum phase are subsequently engulfed by the final BH,
ensuring that the metastable EW vacuum remains observationally unperturbed to the external universe.

The mechanism of EW vacuum turbulence induced during BBH coalescence renders such mergers as highly
transient yet powerful sources of UHE neutrinos. These sources are distributed over redshift, with distances
inferred from gravitational wave observations by LIGO, Virgo, and KAGRA.
To be viable, the spectral and luminosity characteristics of these sources population must simultaneously
meet several constraints: they must support a steady-state detection rate of \(\gtrsim 100\)~PeV neutrinos that remains
consistent with IceCube's non-observation of diffuse UHE neutrinos, while also allowing for the rare detection by IceCube of
events at \(\lesssim 10\)~PeV energies~\cite{IceCube2PeV1,IceCube5PeV1,IceCube6PeV1,IceCube13PeV1}. Furthermore, the model must ensure a non-negligible probability that a single,
exceptionally luminous event is detected if it occurs within KM3NeT/ARCA's field of view.
We demonstrate that a Pareto-type luminosity distribution, when combined with a source spectrum that rises with
neutrino energy, naturally accommodates these constraints.
This framework provides a compelling and flexible emission model capable of reconciling both the lack of diffuse neutrino background and the detection of rare, high-energy neutrino events.

In addition to neutrinos, BBH mergers are expected to produce gamma rays with a spectrum mirroring that of the neutrino emission. As these high-energy gamma rays escape the mergers, they interact with the EBL and the cosmic microwave background (CMB), initiating cosmological electromagnetic cascades.
The resulting secondary gamma-ray spectrum may be detectable by current and upcoming observatories such as LHAASO, HAWC, MAGIC, H.E.S.S., Fermi-LAT and CTA. In this study, we evaluate the predicted detection rates and spectral characteristics of such cascades, using the same Pareto-distributed luminosity function and redshift evolution parameters that were employed to reconcile the IceCube--KM3NeT tension in their joint neutrino detection framework.
We demonstrate that both the steady-state gamma-ray flux arising from the cumulative cascade of the BBH source population and the signal from a single high-luminosity transient source are consistent with the absence of correlated to the
KM3NeT's event electromagnetic counterparts observed by the current gamma ray facilities.
In particular, we show that the predicted steady-state flux in the GeV–PeV range agrees with LHAASO, HAWC and Fermi-LAT measurements, assuming physically motivated spectral indices and luminosity distributions. Additionally, we discuss the potential detectability of short-timescale, burst-like signatures from bright individual BBH mergers with
current and next-generation gamma-ray observatories. We further examine the low energy electromagnetic counterparts of BBH mergers residing in the vicinity of galactic nuclei, emphasizing how reprocessing in different nuclear environments shapes the observable X-ray and ultraviolet (UV)/optical signals.

Finally, it is worth emphasizing that, in the context bubble collision driving turbulence in EW vacuum on instant timescales, BBH mergers emerge as promising multi-messenger phenomena. They are expected to simultaneously emit gravitational waves, UHE neutrinos, and gamma rays, with the latter observable across a broad spectral band after undergoing electromagnetic cascading.

This paper is structured as follows. In Section~\ref{mechanism}, we outline the key features of the underlying mechanism involving gravitational corrections to vacuum decay within the narrow gap between two closely orbiting horizons, as initially introduced and explored in detail in~\cite{tribo}. We also present preliminary estimates for the energy and temporal characteristics of
neutrino emission from \(\mu\)BHs.
Section~\ref{distr} establishes the emission properties required from a population of BBH mergers to reconcile
the UHE neutrino detection by KM3NeT’s ARCA detector with the absence of similar events in IceCube’s decade-long dataset.
In Section~\ref{gamma} (four subsections) we comprehensively quantify the electromagnetic cascades initiated by UHE gamma rays coproduced with neutrinos and assess how reprocessing of these UHE gamma rays proceeds when the merger occurs in different galactic–nucleus environments. In Section~\ref{sec:discussion}, we examine possible roles of percolation in regions where the EW vacuum may
become destabilized, focusing on their potential relevance to the luminosity and spectral features of BBH mergers as
UHE neutrino sources. Finally, we present our conclusions in Section~\ref{sec:summary}.

\section{Higgs field induced neutrino emission from a BBH merger}
\label{mechanism}
When the two components of a stellar-mass BBH system spiral to within a fraction of a Schwarzschild radius of one another, the region that separates their horizons resembles a thin unstable vacuum ``sandwich'' (UVS)~\cite{tribo}.  In that narrow gap the gravitational potentials of the companions almost cancel, so the local space-time is nearly flat while the curvature gradients remain strong.  Those gradients distort the SM Higgs potential, lowering the height of the metastability barrier and making electroweak vacuum decay much more likely.

To estimate a vacuum-decay rate near a BH horizon a toy model was proposed in our early work ~\cite{tribo} based on the effective potential
\begin{equation}
U(h)\;\simeq\;\frac{\lambda}{4}\,h^{4}\!\left(1-\frac{2}{k}\ln\frac{h}{v}\right),
\label{eq:U}
\end{equation}
where the self-coupling constant $\lambda\!\simeq\!0.127$ and the vacuum-expectation value $v = 246\;\text{GeV}$.
Parameter $k$ in Eq.\,\eqref{eq:U} is given by
\begin{equation}
k\;\simeq\;9.3\,\frac{V(R)}{A(R)^{3}}\;\simeq\;0.073\,d_{H}\;\ll 1,
\label{eq:k}
\end{equation}
where $d_{H}$ is the distance to the horizon measured in units of the
Schwarzschild radius $R_{S}$ of the BH.  The metric coefficients are
\begin{equation}
V(R)=\frac{1-\dfrac{M}{2R}}{1+\dfrac{M}{2R}},\qquad
A(R)=\left(1+\frac{M}{2R}\right)^{2},
\label{eq:VA}
\end{equation}
with the radial function $R$ expressed in terms of the Schwarzschild radial coordinate as
\begin{equation}
2R \;=\; r - M + \sqrt{r\bigl(r-2M\bigr)}.
\label{eq:Rofr}
\end{equation}

It follows from~\eq~\ref{eq:U} that in Minkowski spacetime, where V = A = 1, the SM vacuum at h = 0 is meta-stable but the probability of its decay is extremely low. However, in a strong gravitational field close to a BH horizon the parameter k (see~\eq~\ref{eq:k}) decreases and at some distance to the horizon the potential~\eq~\ref{eq:U} becomes negative already at $h \simeq v$ leading to a significant vacuum instability.

With the potential~\eq~\ref{eq:U} the probability of vacuum decay has been calculated by evaluating the four-dimensional bounce action~\cite{tribo}, and reviewed that the decay rate per unit four-volume reads
\begin{equation}
\Gamma(d_{H})\;=\;M^{4}\,
\exp\!\bigl[-480\,d_{H}^{2}\bigr]\ ,
\label{eq:Gamma0}
\end{equation}
where the mass scale \(M\simeq v\) roughly corresponds to the EW vacuum-expectation value.
Therefore, the transition rate can be substantially increased
in the vicinity of the horizon of a BH's horizon so that one expects a high probability
of formation of a bubble of the true Higgs field vacuum.
Nucleation begins with bubbles whose radii exceed the critical value~\cite{tribo}
\begin{equation}
\rho_{c}\;\simeq\;0.045\, v^{-1}(\lambda)^{-1/2}\,
                \sqrt{\frac{k}{0.01}}
          \;\simeq\;10^{-17}\,
                \sqrt{\frac{k}{0.01}}\;\text{cm}\ .
\label{eq:rhoc}
\end{equation}
Once a bubble reaches $\rho_{c}$ it accelerates outward at essentially the speed of light.
In isolated BHs, such bubbles typically fall beneath the event horizon before they can grow
significantly, precluding a catastrophic vacuum transition~\cite{smInstable1}.

In a BBH merger, however, a temporary UVS volume may form between the horizons. For approximately equal-mass BH components, the UVS volume is~\cite{tribo}
\begin{equation}
V_{\rm UVS}(d_H) \approx \frac{5}{2} \pi d_H^2 R_S^3.
\end{equation}
This low-curvature zone allows bubbles to grow before being absorbed. Collisions between bubbles in this region can produce \(\mu\)BHs through double~\cite{pbhBubble4} or triple~\cite{pbhBubble3} wall collisions, which rapidly evaporate via Hawking radiation~\cite{Hawking}.

The masses of such ``split off'' $\mu$BHs are defined by sizes of the bubbles at the instance of
their collision and their vacuum wall tensions.
The surface tension of the bubble walls is estimated as
\begin{equation}
\label{eq:sigmaWall}
\sigma \simeq \sqrt{\lambda} h_b^3~,
\end{equation}
where $h_b \lesssim 1\,\mathrm{TeV}$ is the bounce field value, so that a $\mu$BH of mass
$M_{\mu{\rm BH}}$ is created out of a bubble of radius
\begin{equation}
\label{eq:Rbubble}
R_{\rm bub} \simeq \frac{M_{\mu{\rm BH}}^{1/2}}{(4\pi)^{1/2} \lambda^{1/4} h_b^{3/2} \kappa_f^{1/2}}~,
\end{equation}
with $\kappa_f$ being the fraction of the bubble energy converted to $\mu$BH mass.

Within the volume \(V_{\rm UVS}(d_{\rm H})\), bubble collisions occur with high probability, potentially creating a large number of $\mu$BHs, depending on the nucleation rate $\Gamma(d_{\rm H})$, bubble radius,
and percolation dynamics. In general, there are two scenarios~\cite{tribo}: sparse nucleation leading to isolated bubble collisions; and dense nucleation where bubbles percolate and most undergo double or triple collisions, enabling efficient $\mu$BH formation. Both regimes are likely realized at different stages of the BBH merger. Bellow we
consider second scenario, where the probability of nucleation of a bubble of critical radius $\rho_{c}$ (critical volume \(V_c\)) within the collision time  $\Delta t_{\text{col}}$ should be close to unity. That requirement is expressed by
\begin{equation}
\bigl\langle\Gamma(d_{H})\bigr\rangle\,
\Delta t_{\text{col}}\,
V_{c}\;\simeq\;1\ .
\qquad
%V_{c}=\frac{4\pi}{3}\,\rho_{c}^{3},
\label{eq:percolation}
\end{equation}
This leads to the constraint~\cite{tribo}
\begin{equation}
d_H \lesssim 0.23,
\label{eq:dHbound}
\end{equation}
implying that whenever the dimensionless gap $d_{H}$ lies below roughly $0.23$, bubbles nucleate so profusely that essentially every one of them participates in double and triple wall collisions, guaranteeing abundant $\mu$BH production.

These \(\mu\)BHs are expected to evaporate rapidly via Hawking radiation, emitting Standard Model particles with a nearly thermal spectrum~\cite{pbhEvap}. The differential emission rate for particles of spin \(s\) is given by
\begin{equation}
\frac{d^2N}{dE\,dt} = \frac{\Gamma_s}{2\pi} \left[ \exp\left( \frac{E}{T_{\mu{\rm BH}}} \right) - (-1)^{2s} \right]^{-1},
\label{bhSP1}
\end{equation}
where \(\Gamma_s\) is the spin-dependent greybody factor that accounts for quantum absorption probabilities, and \(T_{\mu{\rm BH}}\) is the temperature of the evaporating \(\mu{\rm BH}\).

The temperature of a \(\mu\)BH of mass \(M_{\mu{\rm BH}}\) is given by
\begin{equation}
T_{\mu{\rm BH}} = \frac{1}{8\pi} M_{\rm Pl} \left( \frac{M_{\rm Pl}}{M_{\mu{\rm BH}}} \right)
\approx 10^2 \left( \frac{10^5\,\mathrm{g}}{M_{\mu{\rm BH}}} \right)\,\mathrm{PeV},
\label{bhT1}
\end{equation}
where \(M_{\rm Pl}\) is the Planck mass. This sets the characteristic energy scale of emitted particles.

The average energies of neutrinos and photons produced by \(\mu\)BH evaporation are determined by the peak of the emission spectrum, with
\begin{equation}
E_\nu \approx 5.1\, T_{\mu{\rm BH}}, \qquad E_\gamma \approx 2.8\, T_{\mu{\rm BH}},
\end{equation}
for fermionic and bosonic species, respectively~\cite{evap1}. These values are consistent with the high-energy regime required to explain events such as those observed by KM3NeT and IceCube.

Particles emitted near the horizon of a BH involved in a BBH
merger experience energy loss as a result of gravitational redshift while
escaping the intense gravitational field.
A particle emitted at radius \( r_{\rm em} = R_{\rm S}(1 + d_{\rm H}) \)
with local energy \( E_{\rm em} \) will be observed at infinity with energy:
\begin{equation} \label{zrgav1}
E_{\rm obs} = \frac{E_{\rm em}}{1 + z_{\rm g}},
\end{equation}
where the gravitational redshift factor \( z_{\rm g} \) is determined by the Schwarzschild metric as~\cite{gravMTW}:
\begin{equation} \label{zgrav2}
z_{\rm g} = \frac{1}{\sqrt{1 - R_{\rm S}/r_{\rm em}}} - 1.
\end{equation}
In terms of the dimensionless offset \( d_{\rm H} \), this simplifies to:
\begin{equation} \label{zgrav3}
1 + z_{\rm g} = \sqrt{ \frac{1 + d_{\rm H}}{d_{\rm H}} }.
\end{equation}
For a representative value \( d_{\rm H} \simeq 0.23 \) (see Eq.~\ref{eq:dHbound}),
this redshift factor implies that the observed energy of neutrinos or other particles
is roughly halved relative to their emission energy, due to propagation through the
intense gravitational potential near the horizon of a component of BBH merger.

The Hawking luminosity of a \( \mu\mathrm{BH} \) is given by
\begin{equation} \label{bhL1}
L_{\mu\rm BH} \approx 3.6\times 10^{29} \left( \frac{10^{8}\,\mathrm{g}}{M_{\mu\rm BH}} \right)^2\ \mathrm{erg/s},
\end{equation}
indicating that the total energy released over its lifetime is approximately
\begin{equation} \label{bhE1}
E_{\mu\rm BH} \approx 3.6\times 10^{25} \left( \frac{M_{\mu\rm BH}}{10^{5}\,\mathrm{g}} \right)\ \mathrm{erg}.
\end{equation}
The corresponding evaporation time is
\begin{equation} \label{bht1}
t_{\rm ev} = \frac{5120\pi}{M_{\rm Pl}} \left( \frac{M_{\mu\rm BH}}{M_{\rm Pl}} \right)^3 \approx 8.4 \times 10^{-11} \left( \frac{M_{\mu\rm BH}}{10^{5}\,\mathrm{g}} \right)^3\ \mathrm{s}.
\end{equation}

In the next section, we adopt a neutrino emission spectrum from BBH mergers that rises with energy and is modeled as a power law, valid at least over a defined energy range from  \(E_{\text{min}}\) to \(E_{\text{max}}\), where the BBH mergers serving
as neutrino sources distributed out to redshift \(z \lesssim 3\). At this stage, it is instructive to provide baseline estimates of the relevant parameters characterizing such mergers as high-energy neutrino sources, particularly in light of the highest-energy neutrino events observed to date: a \( 13\,\mathrm{PeV}\) event reported by IceCube~\cite{IceCubePeV2} and the \( 220\,\mathrm{PeV}\) KM3-230213A event detected by KM3NeT/ARCA~\cite{KM3NeT_220PeV}.

From Eqs.~\ref{bhT1}, \ref{bhE1}, and \ref{bht1}, we infer that generating neutrinos with externally observed energies
around \(E_{\text{min}}=50\,\mathrm{PeV}\), originating from the vicinity of a BBH merger, requires a \(\mu\mathrm{BH}\) of mass approximately
\(500\,\mathrm{kg}\). A \(\mu\mathrm{BH}\) of this mass, emitting neutrinos of \(\sim 50\,\mathrm{PeV}\) after redshifting, could potentially account for events comparable in energy to the highest-energy neutrinos of about \(13\,\mathrm{PeV}\) detected by IceCube~\cite{IceCubePeV2}
from sources residing up to redshift \(z\simeq 3\).
Such a \(\mu\mathrm{BH}\)  would initially emit neutrinos peaking at \(\sim 100\,\mathrm{PeV}\), which are subsequently redshifted by a factor of two due to the strong gravitational field at the emission site (see Eq.~\ref{zrgav1}). The evaporation timescale of a \(500\,\mathrm{kg}\) \(\mu\mathrm{BH}\) is approximately \(10\,\mathrm{ns}\), during which it releases a total energy of \(\sim 2 \times 10^{26}\,\mathrm{erg}\), corresponding to the emission of approximately \(4\times 10^{20}\) neutrinos with initial energies near \(100\,\mathrm{PeV}\).

In the case of observing an event with an energy around \(200\,\mathrm{PeV}\) arriving from a similar redshift (\(z\simeq 3\)), the required source would instead be a \(\mu\mathrm{BH}\) of mass \(\sim 30\,\mathrm{kg}\), emitting neutrinos with initial spectral peak energies of \(\sim 2\,\mathrm{EeV}\), which is then gravitationally red shifted down to \(E_{\text{max}}=1\,\mathrm{EeV}\). Such a \(\mu\mathrm{BH}\) would release a total energy of \(\sim 10^{25}\,\mathrm{erg}\) over an even shorter evaporation timescale (\(\sim 3\,\mathrm{ps}\)), producing a neutrino multiplicity of approximately \(4\times 10^{18}\).

Thus, each BBH merger can serve as a transient host of multiple \(\mu\mathrm{BH}\)s, instantaneous but powerful emitters of
UHE neutrinos, potentially detectable by instruments such as IceCube and KM3NeT.
The detection rate of such neutrinos is governed by the mass distribution of \(\mu\mathrm{BH}\)s
produced in each merger, their overall multiplicity, and the cosmological distribution of BBH
mergers across redshift. In the following section, we examine numerically these characteristics
in the context of BBHs as sources of extremely high-energy neutrinos, with particular attention to
the KM3NeT event KM3-230213A.

\section{Modeling source population based on the KM3NeT's neutrino detection}
\label{distr}

In our setup, BBH mergers produce $\mu{\rm BHs}$
during the final instances of their coalescence. These $\mu{\rm BHs}$ evaporate instantaneously,
releasing neutrinos with energies substantially exceeding 100 PeV. As a result, BBH mergers
can be regarded as transient sources of UHE neutrinos.

To reconcile the detection of a \(\sim 220\)~PeV neutrino by KM3NeT’s ARCA detector with IceCube’s
decade-long non-detection of similar events, we propose that BBH mergers, distributed up to
redshift \( z_* \), act as instantaneous emitters of \( N_{\nu} \) neutrinos.
These sources must sustain a steady-state \(\gtrsim 100\) PeV neutrino detection rate
that remains consistent with IceCube’s null results while ensuring a sufficiently high
probability that at least one event from a single merger is detected, provided it is
favorably positioned within ARCA’s field of view. Within this framework, each BBH merger
at redshift \(z\) must release neutrinos with rest-frame energies satisfying
\beq
\label{Eem}
E_\text{em} \gtrsim 200(1+z) \, \text{PeV}\ .
\eeq
in order to account for the observed event energy after cosmological redshift.

There is no reason to expect monochromatic mass distribution of $\mu{\rm BHs}$ emerging in
individual BBH mergers, since the dynamics of genuine true Higgs field vacuum percolation in BBH mergers suggest that,
within vast clusters of nucleated bubbles, the larger and therefore less
probable bubbles are less likely to survive collisions with smaller more numerous bubbles
that have not yet grown to comparable sizes. As a result, the formation of higher-mass
\(\mu{\rm BHs}\) requires rare interactions between similarly large bubbles, while
smaller-mass \(\mu{\rm BHs}\) are more likely to form either from collisions between
small bubbles themselves or from the disruption of larger bubbles by smaller ones.
This asymmetry naturally leads to a broad, non-monochromatic mass spectrum of \(\mu{\rm BHs}\)
that is biased toward lower masses.

As a result, the associated neutrino emission from BBH mergers is expected to follow a spectrum
that rises with energy, which we model as a power law:
\begin{equation}
\label{bbhSpectr1}
\frac{dN}{dE} \propto \left( \frac{E}{E_0} \right)^{-\gamma},
\end{equation}
with a spectral index \(\gamma < 1\), reflecting the predominance of lower-mass \(\mu{\rm BHs}\) and the resulting hard, high-energy nature of their neutrino output. For such a power-law spectrum, the fraction of neutrinos with energies above the ARCA detection threshold of \(200(1+z)\,\mathrm{PeV}\) within the energy range \(E_{\text{min}}\) to \(E_{\text{max}}\) is given by:
\begin{equation}
\label{fracE}
f_{\text{ARCA}}(z) = \frac{E_{\text{max}}^{1-\gamma} - \left[200(1+z)\right]^{1-\gamma}}{E_{\text{max}}^{1-\gamma} - E_{\text{min}}^{1-\gamma}}.
\end{equation}

To estimate the total (steady-state) neutrino detection rate from BBH mergers,
we begin by considering their comoving number density, commonly modeled as~\cite{zBBHrate1,zBBHrate2,LVKpopulation1}:
\beq
\label{Rz1}
R(z) = R_0 (1 + z)^\kappa,
\eeq
where \( R_0 \) is the local merger rate at redshift \( z = 0 \), ranging from
\( 17.9 \, \text{Gpc}^{-3} \, \text{yr}^{-1} \) to \( 44 \, \text{Gpc}^{-3} \, \text{yr}^{-1} \),
based on LIGO-Virgo-KAGRA (LVK) observations~\cite{LVKpopulation1}.
The parameter \( \kappa \) describes the redshift evolution, typically estimated within the range
\( 1.5 \leq \kappa \leq 2.7 \)~\cite{zBBHrate2,zBBHrate3}.
The proposed mechanism for neutrino emission from BBH mergers assumes no redshift evolution in neutrino production
per merger. Therefore, the total neutrino detection rate can be expressed by integrating the merger rate
\( R(z) \) (as defined in~\eq~\ref{Rz1}), scaled by the expected number of neutrinos per merger
\( \langle N \rangle \), and the detection probability defined by the detector's
effective area, over redshift up to \( z_{*} \):
\beq
\label{nuRate1}
\lambda_{\text{total}} = \int_0^{z_{*}} R(z) \frac{dV}{dz} \langle N \rangle \frac{A_{\text{eff}}}{4\pi D_L(z)^2} dz\ ,
\eeq
where \( \frac{dV}{dz} \) is the comoving volume element, \( D_L(z) \) is the luminosity distance, and \( A_{\text{eff}} \) represents the detector's effective area. The comoving volume element and luminosity distance are given by:
\beq
\label{comV1}
\frac{dV}{dz} = \frac{4\pi c^3}{H_0^3} \frac{(1 + z)^2}{ I(z)} \left( \int_0^z \frac{1}{I(z')} dz' \right)^2\ ,
\eeq
\beq
\label{lumD1}
D_L(z)=\frac{c}{H_0}(1+z)\int_0^z \frac{1}{I(z')} dz'\ ,
\eeq
where \( c \) is the speed of light, \( I(z) = \sqrt{\Omega_m (1 + z)^3 + \Omega_\Lambda} \),
with \( \Omega_m = 0.31 \), \( \Omega_\Lambda = 0.69 \), and \( H_0 = 70 \, \text{km/s/Mpc} \).

Our aim is to determine a neutrino luminosity distribution for BBH mergers that yields, on average,
one detectable neutrino over a timescale of \(\sim 10\) years with IceCube’s effective area
(\(A_{\text{effIC}}\)), while simultaneously ensuring a high probability (\(\gtrsim 50\%\))
of detecting at least one neutrino from an individual merger using a detector ten times smaller
(\(A_{\text{effARCA}} \approx 0.1 A_{\text{effIC}}\)).
To model this, we employ the Pareto distribution, a heavy-tailed probability distribution
well-suited for capturing the statistical behavior of rare but highly luminous events.
In this context, while most BBH mergers emit relatively modest numbers of neutrinos,
a small subset may produce exceptionally intense bursts.

The Pareto probability density function (PDF) is given by:
\beq
\label{paretoPDF1}
f(N; N_{\text{min}}, \alpha) =
\begin{cases}
\frac{\alpha \cdot N_{\text{min}}^\alpha}{N^{\alpha + 1}} & \text{for } N \geq N_{\text{min}}, \\
0 & \text{otherwise},
\end{cases}
\eeq
which gives the cumulative distribution function (CDF) of $N \geq N_{\text{min}}$ as
\beq
\label{paretoCDF1}
{\cal F} (N \geq N_{\text{min}}) = \left(\frac{N_{\text{min}}}{N}\right)^\alpha \ ,
\eeq
where \( N_{\text{min}} \) is the minimum neutrino emission (scale parameter),
\( \alpha \) is the shape parameter controlling the tail heaviness.
The Pareto distribution has parameters \(\alpha\) (shape) and \(N_{\rm min}\) (minimum emission).
The expectation value of N for a Pareto distribution is
\beq
\label{pareto2}
\langle N \rangle\ = \frac{\alpha N_{\text{min}}}{\alpha - 1}\ ,
\eeq
which is further used for calculating the total detection rate \eq~\ref{nuRate1}.

Using the detection rate per unit redshift per volume $R(z)dV/dz$ and \eq~\ref{paretoCDF1} with $N=N_{th}(z)$
the detection Poisson probability of a neutrino from a single source in time range
$\delta t$ can be express as follows
\beq
\label{single1}
P_{\text{single}} = 1 - \exp\left(-\delta t \int_0^{z_*} R_0 (1+z)^\kappa \frac{dV}{dz}
\left(\frac{N_{\text{min}}}{N_{\text{th}}(z)}\right)^\alpha \, dz \right) \ ,
\eeq
where a source emitting \(N_{\text{th}}(z)\) neutrinos at redshift \(z\) has a  probability
\(P_{\text{ARCA}}(N_{\text{th}}(z))\) of being detected in the small detector (ARCA).
This probability is determined from
from Poisson statistics of detection of at least one neutrino from a merger
at redshift \( z \)
\beq
\label{single2}
P_{\text{ARCA}}(N_{\text{th}}(z)) = 1 - \exp\left( -\frac{N_{\text{th}}(z) A_{\text{effARCA}}f_{\text{ARCA}}(z)}{4\pi D_L(z)^2} \right)\ .
\eeq

For numerical estimates, we integrate up to a redshift cutoff of \( z_* = 3 \), which serves as a practical upper limit for the contribution of BBH mergers to the neutrino flux. This choice is motivated by astrophysical population models indicating that the comoving merger rate density of BBH systems increases with redshift, peaking around \( z \sim 2\text{--}3 \), and subsequently declining at higher redshifts due to the falloff in star formation activity and the delay-time distribution of binary evolution~\cite{zBBHrate2}. Specifically, the merger rate at \( z = 3 \) is estimated to be roughly six times higher than the local value before turning over. Beyond this redshift, both the decreasing source population and the increasing luminosity distance significantly suppress the contribution to the detectable neutrino signal. Additionally, the neutrino detection probability scales inversely with the square of the luminosity distance, which becomes increasingly unfavorable at higher \( z \). Thus, setting \( z_* = 3 \) captures the bulk of the astrophysically motivated signal while avoiding overestimation from poorly constrained high-redshift populations.

We adopt the range of the local BBH merger rate \( R_0 = 18\, - 44\, \text{Gpc}^{-3} \, \text{yr}^{-1} \) and the
range of redshift evolution index \( \kappa =1.5\, -  2.3 \), consistent with recent population synthesis and LVK constraints. For the effective area, we use \( A_{\text{effIC}} \simeq 6000 \, \text{m}^2 \) in the direction of
KM3-230213A, as derived from public IceCube data by the authors of Ref.~\cite{Neronov:YearLongSource1,IceCubeData1}.
The corresponding effective area of the KM3NeT/ARCA detector is estimated to be \( A_{\text{effARCA}} \simeq 400 \, \text{m}^2 \)~\cite{Neronov:YearLongSource1,IceCubeData1}, approximately an order of magnitude smaller than that of IceCube.
The exposure time in~\eq~\ref{single1} is set to 10 years.

Given that IceCube has reported three neutrino events with energies exceeding $5\,\mathrm{PeV}$ each identified in independent data samples using different selection criteria~\cite{IceCube2PeV1,IceCube5PeV1,IceCube6PeV1,IceCube13PeV1}. One of these events reached up to $13\,\mathrm{PeV}$, therefore we adopt a lower bound of $E_{\text{min}} = 50\,\mathrm{PeV}$. This choice corresponds to neutrinos with observed energies $\gtrsim 10\,\mathrm{PeV}$, assuming a redshift $z_* = 3$.
Similarly, we set $E_{\text{max}} = 1\,\mathrm{EeV}$ to accommodate the detection of a $\gtrsim 200\,\mathrm{PeV}$ neutrino in the event KM3-230213A, observed by KM3NeT/ARCA.
\begin{table}[htbp]
\caption{\label{tab:pareto-results}
Optimized parameters of the Pareto-distributed neutrino luminosity function for BBH mergers, determined under the constraint of an average IceCube detection rate of 0.10 events per year over a 10-year exposure. Each row corresponds to a different combination of spectral slope $\gamma$ (governing the neutrino energy spectrum $dN/dE \propto E^{-\gamma}$) and target probability $P_{\text{ARCA}}$ for detecting at least one neutrino from a single BBH merger in KM3NeT/ARCA. The parameters $\alpha$ (Pareto shape) and $N_{\text{min}}$ (Pareto scale) define the distribution of total neutrino yields per merger. Columns $\langle N \rangle$ and $\langle E_\nu \rangle$ give the expected neutrino multiplicity and mean energy, respectively. The ARCA detection probability reflects the likelihood of a detectable event from a favorably located merger. The upper section assumes a threshold emission limit of $N_{\text{th}}(z) \leq 10^{48}$, while the lower section corresponds to a stricter constraint $N_{\text{th}}(z) \leq 10^{47}$.
Parentheses indicate ranges from the minimum (18 Gpc\(^{-3}\) yr\(^{-1}\), \(\kappa=1.5\)) to the maximum (44 Gpc\(^{-3}\) yr\(^{-1}\), \(\kappa=2.7\)) parameter combinations in \eq~\ref{Rz1}.
The rows in the lower section, with detection probabilities given in parentheses, indicate failing scenarios, for illustration.
}
\begin{ruledtabular}
\begin{tabular}{cccccc}
\multicolumn{6}{c}{\textbf{\(N_{\text{th}}(z) \leq 10^{48}\) }} \\
\hline
\(\gamma\) & \(P_{\text{ARCA}}\) [\%] & \(\alpha\) & \(N_{\text{min}}\) & \(\langle N \rangle\) & \(\langle E_\nu \rangle\) [PeV] \\
\hline
0.3 & 50 & 1.20--1.30 & (1.5--3.0)$\times 10^{45}$ & (6.0--12)$\times 10^{45}$ & 550 \\
    & 70 & 1.40--1.50 & (4.0--8.0)$\times 10^{45}$ & (1.6--3.2)$\times 10^{46}$ & 520 \\
    & 90 & 1.75--1.85 & (1.5--3.0)$\times 10^{46}$ & (0.8--1.6)$\times 10^{47}$ & 480 \\
0.4 & 50 & 1.25--1.35 & (2.2--4.4)$\times 10^{45}$ & (6.6--13)$\times 10^{45}$ & 490 \\
    & 70 & 1.45--1.55 & (5.0--10)$\times 10^{45}$ & (1.5--3.0)$\times 10^{46}$ & 460 \\
    & 90 & 1.85--1.95 & (2.0--4.0)$\times 10^{46}$ & (1.0--2.0)$\times 10^{47}$ & 420 \\
0.5 & 50 & 1.30--1.40 & (2.8--5.6)$\times 10^{45}$ & (8.4--17)$\times 10^{45}$ & 450 \\
    & 70 & 1.50--1.60 & (6.5--13)$\times 10^{45}$ & (2.3--4.6)$\times 10^{46}$ & 420 \\
    & 90 & 2.00--2.10 & (3.5--7.0)$\times 10^{46}$ & (1.4--2.8)$\times 10^{47}$ & 380 \\
0.7 & 50 & 1.45--1.55 & (5.5--11)$\times 10^{45}$ & (1.7--3.4)$\times 10^{46}$ & 350 \\
    & 70 & 1.70--1.80 & (1.3--2.6)$\times 10^{46}$ & (4.7--9.4)$\times 10^{46}$ & 320 \\
\hline
\multicolumn{6}{c}{\textbf{\(N_{\text{th}}(z) \leq 10^{47}\) }} \\
\hline
0.3 & 50 & 1.50--1.60 & (0.8--1.6)$\times 10^{46}$ & (2.7--5.4)$\times 10^{46}$ & 550 \\
    & (70) & 1.80--1.90 & (2.5--5.0)$\times 10^{46}$ & (1.0--2.0)$\times 10^{47}$ & 520 \\
0.4 & 50 & 1.55--1.65 & (1.0--2.0)$\times 10^{46}$ & (3.1--6.2)$\times 10^{46}$ & 490 \\
    & (70) & 1.85--1.95 & (3.0--6.0)$\times 10^{46}$ & (1.1--2.2)$\times 10^{47}$ & 460 \\
0.5 & 50 & 1.60--1.70 & (1.5--3.0)$\times 10^{46}$ & (4.8--9.6)$\times 10^{46}$ & 450 \\
    & (70) & 1.90--2.00 & (4.5--9.0)$\times 10^{46}$ & (1.8--3.6)$\times 10^{47}$ & 420 \\
0.7 & (50) & 1.80--1.90 & (3.0--6.0)$\times 10^{46}$ & (1.1--2.2)$\times 10^{47}$ & 350 \\
\end{tabular}
\end{ruledtabular}
\end{table}

Gravitational-wave observations by the LVK collaboration indicate that BBH mergers typically radiate between 3\% and 8\% of the system’s total mass in gravitational waves, with measured total masses ranging from \(18\,M_{\odot}\) to \(142\,M_{\odot}\)~\cite{LVKpopulation1}. Based on this, it is reasonable to assume that the total mass-energy available for Hawking evaporation of \(\mu\mathrm{BH}\)s should not exceed \(\sim M_{\odot}\). This sets a physical upper limit on the total neutrino energy budget per BBH merger, which can be translated into a maximum number of neutrinos emitted with energy \(E_{\nu}\):
\begin{equation}
\label{Nth}
N_\text{th}(z) \simeq 3\times 10^{19} \left( \frac{\langle E_\nu \rangle}{\mathrm{1\ EeV}} \right)^{-1} \left( \frac{M_{\odot}}{10^5\,\mathrm{g}} \right)\sim 10^{48}\left( \frac{\langle E_\nu \rangle}{\mathrm{1\ EeV}} \right)^{-1}\,,
\end{equation}
where the average energy \(\langle E_\nu \rangle\) for spectrum \eq~\ref{bbhSpectr1} is expressed as
\beq
\label{Eav1}
\langle E_\nu \rangle = \frac{\int_{E_{\text{min}}}^{E_{\text{max}}} E \cdot E^{-\gamma} \, dE}
{\int_{E_{\text{min}}}^{E_{\text{max}}} E^{-\gamma} \, dE} = \frac{1-\gamma}{2-\gamma} \cdot \frac{E_{\text{max}}^{2-\gamma} -
E_{\text{min}}^{2-\gamma}}{E_{\text{max}}^{1-\gamma} - E_{\text{min}}^{1-\gamma}}.
\eeq
For the neutrino average energies \(\langle E_\nu \rangle\lesssim 500\,\mathrm{PeV}\), this corresponds to \(N_\text{th}(z) \sim 10^{48}\), which we adopt as a conservative luminosity cap.
To ensure physical plausibility, we impose this cap \(N_\text{th}(z) \leq 10^{48}\) in our emission model. For the smaller detector ARCA, this constraint limits the maximum redshift to \(z_\text{cap} \approx 0.8\). Beyond this redshift, the required neutrino luminosities would exceed the allowed budget, and such mergers are excluded from the detectability analysis.
This cap is applied under the assumption that a detection probability of \(P_{\text{ARCA}}\) corresponds to the emission of \(N = N_\text{th}(z)\) neutrinos, based on Poisson statistics \eq~\ref{single2}.
Under this framework, we explore a grid of parameters \((\gamma, \alpha, N_{\text{min}})\), where the spectral slope \(\gamma\) governs the energy distribution \(dN/dE \propto E^{-\gamma}\), and optimize these to satisfy both an average detection rate of 0.1 yr\(^{-1}\) in IceCube (integrated over the entire energy range, wich means of any energy of the spectral range) and a neutrino of energy $E_{\mathrm{1\nu ARCA}}\gtrsim 200\,\mathrm{PeV}$ from a single-merger with detection probability ranging from 50\% to 90\% in ARCA. Cosmological source rate and flux evolution are computed using the \texttt{Astropy} package~\cite{Astropy2013,Astropy2018}.
The outcomes of this optimization are presented in Table~\ref{tab:pareto-results}.

Several systematic trends emerge from our parameter space exploration. Softer spectra (\(\gamma \gtrsim 0.7\)) require significantly higher values of \(N_{\text{min}}\) to generate enough high-energy neutrinos for detection, as the spectral power shifts toward lower energies near \(\sim 50\,\mathrm{PeV}\). Consequently, these scenarios yield lower mean neutrino energies, typically \(\langle E_\nu \rangle \lesssim 400\,\mathrm{PeV}\).
In contrast, harder spectra (\(\gamma \sim 0.3\)) allow for much lower \(N_{\text{min}}\) values because a larger fraction of the fluence is concentrated in the UHE tail. These models produce higher average neutrino energies, often exceeding \(500\,\mathrm{PeV}\), and are well-suited to explain isolated extreme events such as KM3-230213A.
Introducing a stricter luminosity cap of \(N_{\text{th}} \leq 10^{47}\) imposes severe constraints on viable parameter combinations. It shifts \(N_{\text{min}}\) into the range of \(10^{46} - 10^{47}\), significantly raising the total energy output per merger and potentially conflicting with physical limits on BBH energetics. These effects are most evident in lower section of
Table~\ref{tab:pareto-results}, particularly in the rows with detection probabilities given in parentheses, which reflect failing scenarios.
From these trends, we identify a preferred scenario that balances detectability, physical plausibility, and energetic feasibility. The optimal region lies at spectral slopes of \(\gamma = 0.3\)–0.5, with the cap \(N_{\text{th}} \leq 10^{48}\). As an example, the case with \(\gamma = 0.3\) and \(P_{\text{ARCA}} = 70\%\)
yields \(N_{\text{min}} = (4.0-8.0) \times 10^{45}\) neutrinos per merger and a mean energy \(\langle E_\nu \rangle = 520\,\mathrm{PeV}\).
In contrast, parameter combinations with \(\gamma \geq 0.7\) or a luminosity ceiling of \(N_{\text{th}} \leq 10^{47}\), especially when aiming for high detection probabilities (\(P_{\text{ARCA}} \gtrsim 70\%\)), generally exceed acceptable energy limits for BBH systems.
These optimized emission configurations and their associated diagnostics provide a coherent framework for interpreting rare, UHE neutrino detections as manifestations of BBH mergers undergoing electroweak vacuum instability.

%%%%%%%%%%%%%%%%%%%%%%%%%%%%%%%%%%%%%%%%%%%%%%%%%%%%%%%%%%%%%%%%%%%%%%%%%%%%

\section{The electromagnetic cascade}
\label{gamma}

\subsection{Gamma rays absorption and inverse Compton scattering}
\label{absorpGamma}

Alongside neutrinos, the \(\mu\mathrm{BH}\)s emit gamma rays across an extreme energy range (100-1000 PeV). Escaping
from BBH mergers, these gamma rays initiate electromagnetic cascades via interactions
with the EBL and the CMB (see for review~\cite{Sigl:book,Aharonian:book}). The resulting secondary
gamma-ray spectrum may be observable by current and future instruments such as LHAASO, HAWC, MAGIC, H.E.S.S., Fermi LAT and CTA.
In this section, we quantify the expected detection rates and spectra, employing the same Pareto-distributed
luminosity function and redshift evolution as previously introduced for the neutrino scenario.

The cascade physics for photons with energies \(E_{\gamma} \simeq 100\,\mathrm{PeV}\) and above proceeds primarily through pair production with EBL photons~\cite{gEBL2, gEBL3}, whose energies span \(\epsilon \sim 10^{-2}\,\mathrm{eV} - 10\,\mathrm{eV}\). For gamma rays of energy \(E_\gamma\), the dominant absorption process is pair production (\(\gamma\gamma \rightarrow e^+e^-\)) with EBL photons of characteristic energy~\cite{nikishov, gEBL1}:
\begin{equation}
\label{thr1}
\epsilon_{\text{char}}(E_\gamma) \approx 3.5\frac{(m_ec^2)^2}{E_\gamma}\, ,
%E_{\gamma} \ge \frac{m_e^2}{\epsilon} \simeq 260\left(\frac{\epsilon}{1\,\mathrm{eV}}\right)^{-1}\,\mathrm{GeV}~.
\end{equation}
where the factor 3.5 corresponds to the peak cross-section energy (see bellow).
The mean free path (MFP) for this process is given approximately by
\begin{equation}
\label{thr2}
\lambda_{\gamma\gamma}(E_{\gamma}) \simeq \frac{1}{\sigma_{\gamma\gamma}(E_{\gamma}) n_{\rm EBL}}~,
\end{equation}
where the interaction of gamma rays with the extragalactic background light (EBL) via pair production ($\gamma\gamma \to e^+e^-$) is governed by the Breit-Wheeler cross-section~\cite{berest, QED}:
\begin{equation}
\label{siggg}
\sigma_{\gamma\gamma}(E_{\text{CM}}) = \frac{3}{16} \sigma_T (1 - \beta^2) \left[ 2\beta (\beta^2 - 2) + (3 - \beta^4) \ln \left( \frac{1+\beta}{1-\beta} \right) \right],
\end{equation}
with $\beta = \sqrt{1 - (2m_ec^2/E_{\text{CM}}})^2$, center-of-mass energy
$E_{\mathrm{CM}} = \sqrt{2 E_\gamma \varepsilon_{\text{EBL}} (1-cos\theta)}$ where $\theta$ is a collision angle, and Thomson cross section $\sigma_T = 6.65 \times 10^{-25}\, \text{ cm}^{2}$.
In the Klein-Nishina (KN) regime \(E_{\text{CM}} \gg m_e c^2\), corresponding to the high-energy tail, the cross section can be approximated as

\begin{equation}
\label{sigKN}
\sigma_{\gamma\gamma} \approx 3 \sigma_T \left( \frac{m_e^2 c^4}{E_{CM}^2} \right) \ln \left( \frac{E_{CM}}{m_e c^2} \right)\ .
\end{equation}

\begin{figure}[htbp]
  \centering
  \includegraphics[width=0.8\linewidth]{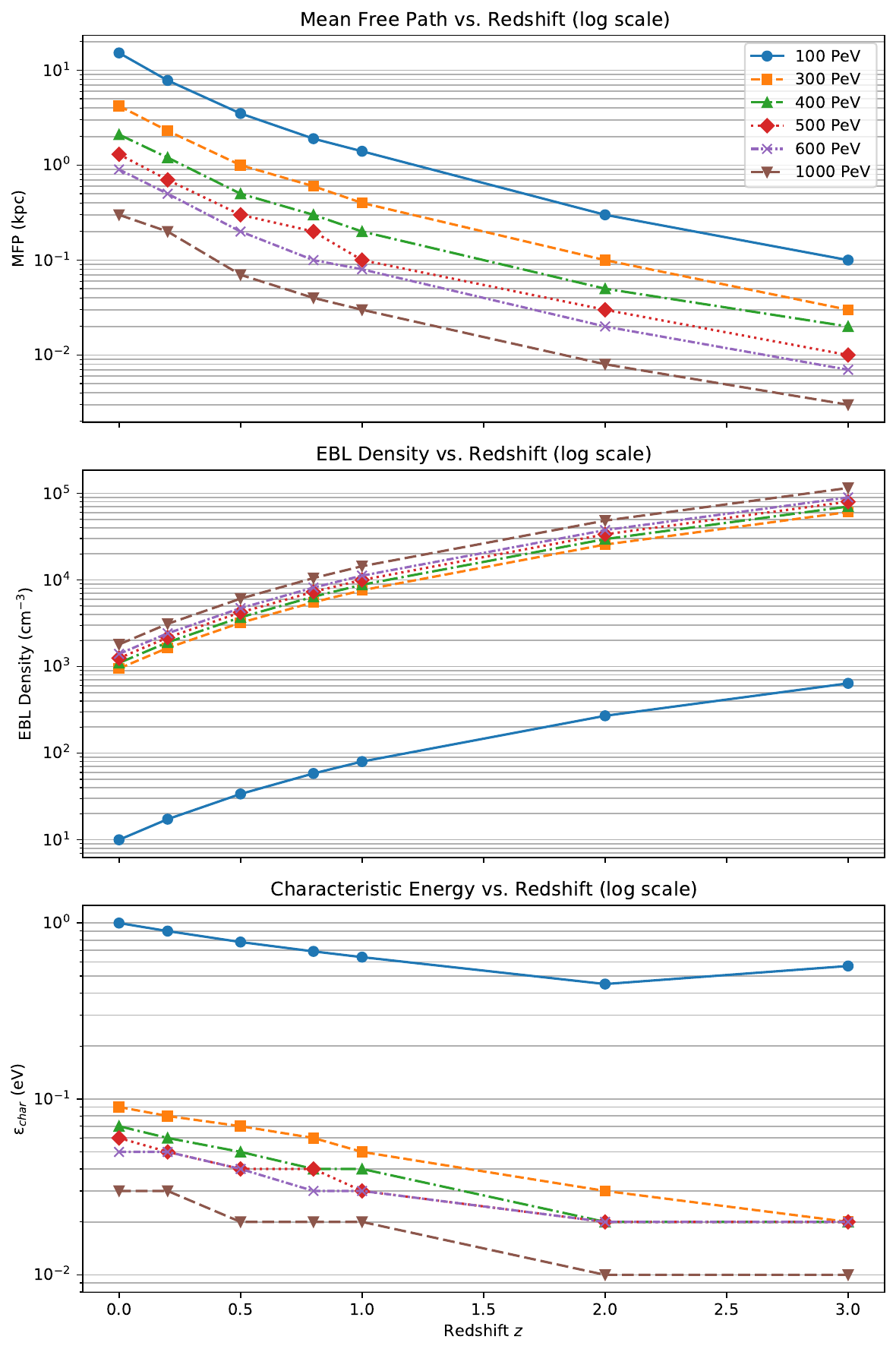}
  \caption{\label{tab:mfp-ebl} MFP of gamma rays due to pair production on the EBL, computed using \textsc{CRPropa~3} with the model of Ref.~\cite{modelEBL1}. The upper panel shows the energy-dependent MFP as a function of redshift. The middle panel presents the corresponding EBL photon number density, while the lower panel displays the characteristic interaction energy \(\epsilon_{\text{char}}\) at which pair production is most efficient.}

  \label{fig:mfp-ebl}
\end{figure}

Gamma rays with energies \(\gtrsim 100\,\mathrm{PeV}\) experience energy-dependent KN suppression in their pair-production cross sections. At \(E_\gamma \sim 100\,\mathrm{PeV}\), interactions occur deep within the KN regime, predominantly involving optical and ultraviolet EBL photons with characteristic energies \(\epsilon_{\mathrm{EBL}} \sim 1\text{--}10\,\mathrm{eV}\). Although the cross section is strongly suppressed \(\sigma / \sigma_T \sim 10^{-5} \text{--} 10^{-6}\), the high number density of target photons compensates, leading to relatively short mean free paths (MFPs) of approximately \(15\,\mathrm{kpc}\) at redshift \(z = 0\). At higher energies, \(E_\gamma \sim 300\text{--}1000\,\mathrm{PeV}\), pair production remains in deep KN regime, with cross section suppression of \(\sigma / \sigma_T \sim 10^{-3} \text{--} 10^{-4}\). In this regime, gamma rays primarily interact with far-infrared (FIR) photons (\(\epsilon_{\mathrm{EBL}} \sim 0.01\text{--}0.03\,\mathrm{eV}\)). While the cross section remains reduced, the elevated FIR photon number density (\(n_{\mathrm{EBL}} \sim 10^3\text{--}10^4\,\mathrm{cm}^{-3}\)) results in very short MFPs—sub-kiloparsec at \(z = 0\), with values as low as \(\sim 0.3\,\mathrm{kpc}\) for \(E_\gamma = 1000\,\mathrm{PeV}\). The redshift evolution of the EBL further enhances this attenuation, as the comoving photon density scales approximately as \(n_{\mathrm{EBL}}(z) \propto (1 + z)^3\), reducing the MFP by over two orders of magnitude out to \(z = 3\).

These effects we quantify using \texttt{CRPropa~3} simulations~\cite{CRPropa} with the EBL model of Ref.~\cite{modelEBL1}, which incorporates the full energy dependence of KN suppression across all regimes. The MFPs were calculated by integrating the Breit--Wheeler cross section~\eq~\ref{siggg} over the evolving EBL spectrum at each redshift. The results, summarized in top panel of Fig.~\ref{tab:mfp-ebl}, reveal two key trends. First, MFPs exhibit strong redshift evolution: for 100~PeV photons, the MFP decreases from \(15.2\,\mathrm{kpc}\) at \(z = 0\) to \(0.1\,\mathrm{kpc}\) at \(z = 3\); for 1000~PeV photons, the MFP drops from \(0.3\,\mathrm{kpc}\) to just \(0.003\,\mathrm{kpc}\) over the same range, reflecting the faster \((1+z)^3\) scaling of FIR photon densities. Second, the characteristic EBL photon energy evolves as \(\epsilon_{\mathrm{EBL}}(z) \propto (1 + z)^{-0.6}\)~\cite{modelEBL3,modelEBL2}~\footnote{Empirical fit shows that stellar population aging and dust reprocessing soften the pure (\(1 + z)^{-1}\) cosmological dilution.}, shifting the dominant interaction partners from optical/UV photons (\(\sim 1\,\mathrm{eV}\)) at \(z = 0\) to FIR photons (\(\sim 0.03\,\mathrm{eV}\)) at \(z = 3\) for 100~PeV gamma rays, and from FIR photons to submillimeter photons (\(\sim 0.01\,\mathrm{eV}\)) for 1000~PeV gamma rays. This energy-dependent behavior implies that gamma rays above \(\sim 300\,\mathrm{PeV}\) are efficiently absorbed over sub-kiloparsec distances and can only be observed from Galactic sources, whereas 100~PeV gamma rays may still probe extragalactic environments out to \(z \lesssim 0.5\).

Consequently, any gamma-ray burst with energy \(\gtrsim 100\,\mathrm{PeV}\) originating from an extragalactic source will be entirely attenuated due to pair production with the EBL. Even for Galactic sources, the flux is expected to be significantly suppressed by a factor \(\exp(-D/\lambda_{\gamma\gamma})\), provided that the MFP \(\lambda_{\gamma\gamma}\) (see middle panel of Fig.~\ref{tab:mfp-ebl}) falls below the typical spatial scale of the Milky Way, in particular for \(E_{\gamma}\) above 300~PeV.
As a result, the energy initially carried by the high-energy \(\gamma\)-rays is efficiently transferred to ultra-relativistic \(e^+e^-\) pairs.

For the energy range of gamma rays from 100~PeV to 1EeV the center-of-mass energy of their interaction with their corresponding characteristic energies of EBL  \(E_{\text{CM}} \gg m_ec^2\).
These secondary electrons and positrons subsequently lose their energy efficiently through inverse Compton scattering on CMB photons, feeding the development of an electromagnetic cascade.
For \(E_e \sim 50\,\mathrm{PeV} - 500\,\mathrm{PeV}\) electrons (with Lorentz factor \(\gamma_e \sim 10^{11} - 10^{12}\)) interacting with CMB photons of energy \(\epsilon_{\text{CMB}} \approx 6.6 \times 10^{-4}\,\mathrm{eV}\), the characteristic IC peak arises in the KN regime. In this limit, the typical energy transfer fraction is
\begin{equation}
\label{fIC0}
f(E_e) \simeq \frac{m_e^{2}c^{4}}{E_e\,\varepsilon_{\mathrm{CMB}}}
           \approx 4\times10^{-3}\left(\frac{E_e}{100~\mathrm{PeV}}\right)^{-1}\ ,
\end{equation}
and assuming the average collision angle $\theta = \pi /2$ the KN cross section is approximately:
\begin{equation}
\label{sigmaKN1}
\sigma_{\text{KN}}(y) \approx \frac{3}{4}\sigma_T y^{-1}\ln(2y)\ ,\quad y \equiv \frac{\gamma_e \epsilon_{\text{CMB}}}{m_e c^2} \simeq 1.3 \times 10^2\left(\frac{E_e}{100\text{PeV}}\right)~.
\end{equation}
The upscattered photon spectrum peaks when the product \(f\sigma_{\text{KN}}(y)\) is maximized~\cite{Aharonian:book}:
\begin{equation}
\frac{dN}{dE} \propto \frac{\ln(2y)}{y^2}~,
\end{equation}
yielding a spectral peak at
\begin{equation}
E_{\text{IC}}^{\text{peak}} \approx \frac{m_e^2 c^4}{\epsilon_{\text{CMB}}} = \frac{(511\,\mathrm{keV})^2}{6.6 \times 10^{-4}\,\mathrm{eV}} \approx 400\,\mathrm{TeV}~.
\end{equation}

While the maximum photon energy in a single IC event may reach
\(E_{\text{IC}}^{\text{max}} \sim \gamma_e m_e c^2 = 50 - 500\,\mathrm{PeV}\),
such events are exponentially suppressed. The 400TeV peak dominates the photon number, whereas the PeV tail dominates the energy budget. The full spectral shape requires numerical integration of the KN cross section~\cite{comptonKN1}.

The mean free path for inverse Compton scattering, \(\lambda_{\text{IC}}\), for electrons with Lorentz factors \(\gamma_e \sim 10^{11}\), is given by
\newcommand{\Eref}{100\,\mathrm{PeV}}
\newcommand{\yref}{1.3\times10^{2}}  % y at 100 PeV for CMB targets
\newcommand{\lambdaref}{38\,\mathrm{kpc}} % λ_IC at 100 PeV, z = 0
\begin{equation}
\lambda_{\mathrm{IC}}(E_e)\;\simeq\;\ \frac{1}{n_{\text{CMB}} \sigma_{\text{KN}}}\;\simeq
\lambdaref\,
\left(\frac{E_e}{\Eref}\right)\,
%\frac{\ln\!\bigl(4\,\yref\bigr)}
\frac{\ln\!\bigl(260)}{\ln\!\bigl[260(E_e/\Eref)\bigr]}\ ,
%     {\ln\!\bigl[4\,\yref\,(E_e/\Eref)\bigr]}\ ,
\label{eq:mfp_scaled}
\end{equation}
where
 \(n_{\text{CMB}} \approx 410\,\mathrm{cm}^{-3}\) is the CMB photon number density at redshift \(z = 0\). Owing to the redshift dependence of the CMB photon density, which scales as \((1 + z)^3\), the mean free path correspondingly decreases by factor 64 at \(z = 3\).

\subsection{Steady state electromagnetic flux}
\label{overalEM}

The upscattered photons which are now the leading particles after the two-step cycle described above still
carry most of the energy of the original gamma ray, and can initiate a fresh cycle of pair production
and IC interactions. This leads to the development of electromagnetic cascades which play an important
role in the resulting observable gamma ray spectra. The cascade evolves through successive
generations of pair production and IC scattering until the photon energy falls below the pair-production
threshold ($\sim 26$\,GeV for optical/UV interactions). At each step,
the energy is partitioned among secondary particles, progressively softening the spectrum.
The bulk of the cascade's energy flux accumulates in the 10–400\, TeV range due to KN suppression at higher energies,
while the photon flux extends down to 10 GeV via late-stage IC and synchrotron emission.

We analyze steady state gamma-ray flux from BBHs that emit gamma rays following the same patterns of the
neutrinos luminosity, assuming the same BBH mergers redshift distribution~\eq~\ref{Rz1} and intrinsic energy
spectrum~\eq~\ref{bbhSpectr1}.
We adopt a fixed gamma-ray to neutrino fluence ratio of \(\Phi_\gamma \simeq 0.6\,\Phi_\nu\) at the source, consistent with the democratic branching of Hawking radiation into Standard Model particles~\cite{pbhEvap,evap1}.
Our analysis considers power-law injection spectra with
spectral indices \(\gamma = 0.3, 0.4, 0.5\), and includes attenuation due to pair production on the EBL and CMB, followed by subsequent electromagnetic cascading. The gamma ray luminosity of the sources follows the same capped Pareto distribution
optimized in Section~\ref{distr} for above spectral indexies.

\begin{table}[htbp]
\caption{\label{tab:flux-gamma-blocks-final} Predicted gamma-ray fluxes from BBH mergers with full uncertainty propagation (68\% confidence interval (CI)). Fluxes are reported in erg\,cm\(^{-2}\)\,s\(^{-1}\)\,sr\(^{-1}\), based on the ranges of \(\alpha\) and \(0.6\times N_{\min}\) optimized for a detection probability \(P_{\mathrm{ARCA}} = 70\%\) (see Table~\ref{tab:pareto-results}) for each corresponding spectral index \(\gamma\). The flux calculations assume a local BBH merger rate \(R_0 = 18\text{--}44\,\mathrm{Gpc}^{-3}\,\mathrm{yr}^{-1}\), redshift evolution parameter \(\kappa = 1.5\text{--}2.7\), and IGMF strength \(B = 10^{-17}\text{--}10^{-14}\,\mathrm{G}\).}
\begin{ruledtabular}
\begin{tabular}{llcc}
\textbf{Spectral Index} & \textbf{Detector (Energy Range)} & \textbf{Observed / Sensitivity$^\dagger$} & \textbf{Predicted Flux (68\% CI)} \\
\hline
\multirow{6}{*}{$\gamma = 0.3$}
 & LHAASO (100\,TeV–1\,PeV) & $10^{-12}$\cite{LHAASO2025} & $(1.8$--$11)\times10^{-13}$ \\
 & HAWC (1–100\,TeV) & $2\times10^{-11}$\cite{HAWC2023} & $(3.6$--$22)\times10^{-12}$ \\
 & H.E.S.S. (1–100\,TeV) & $10^{-11}$\cite{HESS2014} & $(3.0$--$18)\times10^{-12}$ \\
 & MAGIC (0.1–100\,TeV) & $10^{-11}$\cite{MAGIC2020} & $(2.6$--$16)\times10^{-12}$ \\
 & Fermi--LAT (1–100\,GeV) & $10^{-8}$\cite{FERMI2015} & $(0.7$--$1.5)\times10^{-8}$ \\
 & CTA (20\,GeV–300\,TeV) & $10^{-13}$\cite{CTAConsortium2019} & $(5.4$--$41)\times10^{-13}$ \\
\hline
\multirow{6}{*}{$\gamma = 0.4$}
 & LHAASO (100\,TeV–1\,PeV) & $10^{-12}$\cite{LHAASO2025} & $(1.5$--$9.0)\times10^{-13}$ \\
 & HAWC (1–100\,TeV) & $2\times10^{-11}$\cite{HAWC2023} & $(3.0$--$18)\times10^{-12}$ \\
 & H.E.S.S. (1–100\,TeV) & $10^{-11}$\cite{HESS2014} & $(2.5$--$15)\times10^{-12}$ \\
 & MAGIC (0.1–100\,TeV) & $10^{-11}$\cite{MAGIC2020} & $(2.2$--$13)\times10^{-12}$ \\
 & Fermi--LAT (1–100\,GeV) & $10^{-8}$\cite{FERMI2015} & $(0.6$--$1.2)\times10^{-8}$ \\
 & CTA (20\,GeV–300\,TeV) & $10^{-13}$\cite{CTAConsortium2019} & $(4.5$--$34)\times10^{-13}$ \\
\hline
\multirow{6}{*}{$\gamma = 0.5$}
 & LHAASO (100\,TeV–1\,PeV) & $10^{-12}$\cite{LHAASO2025} & $(1.2$--$7.2)\times10^{-13}$ \\
 & HAWC (1–100\,TeV) & $2\times10^{-11}$\cite{HAWC2023} & $(2.4$--$15)\times10^{-12}$ \\
 & H.E.S.S. (1–100\,TeV) & $10^{-11}$\cite{HESS2014} & $(2.0$--$12)\times10^{-12}$ \\
 & MAGIC (0.1–100\,TeV) & $10^{-11}$\cite{MAGIC2020} & $(1.8$--$11)\times10^{-12}$ \\
 & Fermi--LAT (1–100\,GeV) & $10^{-8}$\cite{FERMI2015} & $(0.5$--$1.0)\times10^{-8}$ \\
 & CTA (20\,GeV–300\,TeV) & $10^{-13}$\cite{CTAConsortium2019} & $(3.8$--$28)\times10^{-13}$ \\
\end{tabular}
\end{ruledtabular}
\begin{flushleft}
$^{\dagger}$ Projected sensitivities for CTA; all other values represent published diffuse or isotropic flux measurements averaged over the sky region quoted in the respective reference.
\end{flushleft}
\end{table}

The predicted gamma-ray fluxes in Table~\ref{tab:flux-gamma-blocks-final} are presented as 68\% confidence intervals to properly account for systematic uncertainties across the complete modeling chain, from source emission to detector response. The computational framework centers on \texttt{CRPropa 3}~\cite{CRPropa}, which handles the core physics of gamma-ray propagation through intergalactic space. This Monte Carlo simulator tracks the complete trajectory of photons and particles, modeling three critical processes: pair production via $\gamma + \gamma_{\text{EBL}} \rightarrow e^+ + e^-$ interactions, inverse Compton scattering of secondary $e^\pm$ pairs, and synchrotron radiation in IGMF environments. The simulations incorporate energy-dependent interaction lengths and full 3D particle tracking, with modular components allowing systematic variation of physical parameters. Cosmological quantities such as comoving volumes and luminosity distances were computed using the \texttt{Astropy} package~\cite{Astropy2013,Astropy2018}. To model detection rates, simulated gamma-ray spectra were folded with the instrument response functions using \texttt{Gammapy}~\cite{Gammapy2023}.

Source-intrinsic uncertainties dominate the variance in the predicted gamma-ray fluxes, primarily due to the underlying Pareto-distributed luminosity with shape parameter \(\alpha\) and minimum emission threshold \(N_{\min}\) per merger (see Table~\ref{tab:pareto-results}). For uncertainty propagation, we adopt spectral index–dependent ranges of \(\alpha\) and scale the associated \(N_{\min}\) values by a factor of 0.6, consistent with the assumed neutrino-to-gamma-ray emission ratio. These values we took corresponding to those inferred from the neutrino luminosity function for a KM3NeT/ARCA detection probability of \(P_{\mathrm{ARCA}} = 70\%\) (see Table~\ref{tab:pareto-results}).
Lower values of \(N_{\min}\) for harder injection spectra (e.g., \(\gamma = 0.3\)) imply that fewer low-luminosity sources are needed to account for rare, high-flux events such as KM3-230213A. In contrast, softer spectra (e.g., \(\gamma = 0.5\)) require higher \(N_{\min}\) values to offset the suppressed production of high-energy gamma rays. These input parameters define the distribution of source luminosities from which we generate \(10^4\) Monte Carlo realizations per spectral index \(\gamma\) to compute the 68\% confidence intervals reported in Table~\ref{tab:flux-gamma-blocks-final}. The resulting total gamma-ray output does not exceed \(6 \times 10^{47}\) photons per BBH merger, reflecting the 0.6 scaling of the neutrino luminosity cap used in Table~\ref{tab:pareto-results}.

\begin{table}[htbp]
\caption{\label{tab:uncertainty} Uncertainty contributions for $\gamma=0.4$ to gamma-ray flux predictions ($P_{\text{ARCA}}=70\%$).}
\begin{ruledtabular}
\begin{tabular}{lcc}
\textbf{Source of Uncertainty} & \textbf{Magnitude} & \textbf{Notes} \\
\hline
Luminosity Distribution & $\pm25\%$ & Dominated by $\alpha = 1.45$--$1.55$ and $N_{\min} = (3.0$--$6.0)\times10^{45}$ \\
EBL Model               & $\pm25\%$ & \cite{modelEBL2} (fiducial) vs. \cite{modelEBL1} at $z \leq 3$ \\
IGMF Strength           & $\pm20\%$ & $B = 10^{-17}$--$10^{-14}$ G (cascade suppression) \\
Detector Effective Area & $\pm10\%$ & Energy-dependent (CTA: $\pm8\%$, Fermi-LAT: $\pm12\%$) \\
Spectral Index $\gamma$ & $\pm5\%$  & $\gamma = 0.38$--$0.42$ around fiducial 0.4 \\
Merger Rate $(R_0, \kappa)$ & $\pm15\%$ & $R_0 = 18$--$44$ Gpc$^{-3}$\,yr$^{-1}$, $\kappa = 1.5$--$2.7$ \\
\hline
Total Uncertainty       & $\pm43\%$ & Quadrature sum \\
\end{tabular}
\end{ruledtabular}
\end{table}

Propagation uncertainties contribute comparably to the total variance, with EBL model
uncertainties, which is~\cite{modelEBL2} (fiducial) versus~\cite{modelEBL1}
leading to approximately $\pm30\%$ variations in optical depth $\tau(E,z)$. The IGMF strength,
spanning $10^{-17}$--$10^{-14}$\,G in our simulations, modulates cascade development efficiency by $20\%$--$50\%$
through its influence on $e^\pm$ pair propagation,
leading to a net \(20\%\)--\(30\%\) variation. These effects are rigorously modeled in \texttt{CRPropa~3}.
Detector systematics introduce energy-dependent effects, primarily through uncertainties in the effective area (for example \(\pm15\%\)
 for LHAASO at 100\,TeV and \(\pm10\%\) for Fermi-LAT at 1\,GeV), as well as through limitations in energy resolution.
 Like the optimization of Pareto distribution parameters carried out for the neutrino flux (see Table~\ref{tab:pareto-results}),
 the estimation of the electromagnetic flux incorporates additional cosmological uncertainties associated with the BBH merger rate,
 specifically \(R_0 = 18-44\,\mathrm{Gpc}^{-3}\,\mathrm{yr}^{-1}\)
 and an evolutionary index \(\kappa = 1.5\)--2.7.

The complete Monte Carlo workflow propagates uncertainties through five key stages: (1) Pareto-distributed luminosity sampling, (2) spectral generation for each $\gamma$ value, (3) EBL attenuation with model variations, (4) cascade simulation across IGMF strengths, and (5) convolution with detector responses. As detailed in Table~\ref{tab:uncertainty}, the quadrature sum of individual uncertainties yields $\pm 43\%$ total variation for spectral index \(\gamma =0.4\) and Fermi-LAT as a detector example, with the 16th--84th percentile ranges of $10^4$ realizations defining our reported flux intervals.

Among the spectral indices considered, \(\gamma = 0.3\) yields gamma-ray fluxes that are broadly consistent with observations from \textsc{LHAASO} and \textsc{HAWC}. This harder injection spectrum, however, also leads to a more prominent electromagnetic cascade in the 1--100\,GeV range, contributing to a diffuse component that can approach or marginally exceed the isotropic gamma-ray background (IGRB) observed by \textit{Fermi}--LAT. To remain within IGRB constraints, the total contribution from BBH mergers must not exceed \(\lesssim 10^{-8}\)\,erg\,cm\(^{-2}\)\,s\(^{-1}\)\,sr\(^{-1}\)\,\footnote{Typically quoted with \(\pm 30\%\) systematics~\cite{FERMI2015}.}. Scenarios with \(\gamma = 0.3\) may still remain viable if additional attenuation mechanisms, such as stronger IGMFs (\(B \gtrsim 10^{-12}\)\,G), act to suppress the cascade flux in the GeV band.
The softer spectral index \(\gamma = 0.5\) results in gamma-ray fluxes that stay comfortably below detection thresholds of instruments like \textsc{LHAASO}, \textsc{MAGIC}, and in agreement with IGRB observed
by\textit{Fermi}--LAT. While this avoids tension with existing gamma-ray measurements, it tends to underproduce TeV--PeV gamma-ray fluxes unless BBH merger rates are pushed close to the upper end of current gravitational-wave constraints.
An intermediate case, \(\gamma = 0.4\), achieves a good overall consistency with both neutrino and gamma-ray observations. It allows for flux levels that match \textsc{LHAASO} and \textsc{HAWC} detections without overshooting the IGRB constraint from \textit{Fermi}--LAT and supports the production of UHE neutrinos such as the \(\sim 220\)\,PeV event observed by \textsc{KM3NeT}/ARCA. This spectral index also remains compatible with a capped neutrino luminosity distribution with \(N_{\mathrm{th}} \leq 10^{48}\)\,erg, without necessitating unrealistically high BBH source densities.
Overall, \(\gamma = 0.3\text{--}0.5\) spans a range that is broadly compatible with current multi-messenger constraints. Within this range, \(\gamma = 0.4\) provides a favorable balance between efficient UHE neutrino production and control over the gamma-ray cascade. Nearby (\(z \lesssim 0.2\)) sources remain most strongly constrained by \textsc{LHAASO}, while the upcoming \textsc{CTA} observatory will extend coverage to fainter and more distant mergers.

Future improvements will leverage JWST measurements for EBL constraints, CTA polarization studies for IGMF characterization, and next-generation gravitational wave detectors for merger rate refinement. However, the current implementation already provides robust predictions, with at least $10^{-13}$--$10^{-12}$\,erg\,cm$^{-2}$\,s$^{-1}$\,sr$^{-1}$  flux range for $\gamma=0.4$ remaining statistically significant across all uncertainty realizations. The combination of \texttt{CRPropa}'s detailed physics modeling with comprehensive uncertainty propagation ensures these predictions are both reliable and testable against current and future gamma-ray observations.

\subsection{Single source electromagnetic signature}
\label{singleEM}

Here we consider signal of a single BBH source with cap luminosity of $6\times 10^{47}$ gamma rays with spectral index
$\gamma = 0.4$ which would appear to be favorably exposed to the field of view of gamma ray facilities discussed above.
The predicted fluxes along with respective detection significances are evaluated for the source placed at
redshifts $z = 0.1, 0.2, 0.5, 0.8$. The choice of the maximal redshift
is motivated by the fact that the neutrino cap $N_{\nu}=10^{48}$ to be detected in the KM3NeT/ARCA
limits the maximum redshift to \(z_\text{cap} \approx 0.8\).
The primary gamma rays of energy $\ge $50\,PeV are fully attenuated, producing secondary
GeV--TeV emission through electromagnetic cascades.
The cascaded flux is calculated through the integral:
\begin{equation}
\label{singleFl1}
F_{\text{casc}}\approx \frac{0.5}{4\pi D_L^2\mathcal (z+1)^{4}} \int_{E_1}^{E_2} \left( \frac{dN}{dE} \cdot E \cdot f_{\text{IC}}(E, z) \right) dE,
\end{equation}
where $f_{\text{IC}}(E, z)$ represents the energy- and redshift-dependent fraction of primary energy converted to detectable IC photons
in an energy band relevant for a given detector.
The analysis assumes a total gamma-ray yield of
$N_\gamma = 10^{48}$ with spectrum $dN/dE \propto E^{-0.4}$, where 50\% (factor 0.5 in~\eq~\ref{singleFl1})
of the primary energy
converts to IC gamma rays~\footnote{The apparent 50\% efficiency limitation arises from cascade physics rather than simple energy partitioning. During the first IC scattering event, an electron, emerged from \(\gtrsim 100\)\,PeV primary gama rays, upscatters CMB photons to $\gtrsim$50\,PeV, but these immediately pair-produce on the EBL/CMB. This process inherently partitions half the energy into observable IC gamma rays (GeV--TeV) while the remainder remains in $e^\pm$ kinetic energy for subsequent interactions. Additional energy losses occur through undetectable low-energy photons and adiabatic cosmic expansion.} in weak IGMF ($B = 10^{-15}$\,G). The field of view efficiency varies by detector: LHAASO/KM2A and HAWC cover approximately 2\,sr, CTA observes 0.2\,sr in pointed mode, while Fermi-LAT monitors 2.4\,sr of the sky.

The energy content of the primary gamma rays flash originating from the BBH source implies that
the source release its energy within about 10~ns (see Section~\ref{mechanism} for details), however since electrons
have non-zero rest mass, the secondary gamma rays produced via IC scattering
experience a time delay relative to the primary emission. The delay is approximately given by:
\begin{equation}
\Delta t(E_e) \simeq \frac{D_{\rm IC}}{c} \cdot \frac{m_e^2}{2E_e^2}~,
\label{eq:time_delay}
\end{equation}
where \(D_{\rm IC}\) is the characteristic Compton cooling distance, \(E_e\)
is the energy of the scattering electron, \(m_e\) is the electron mass, and \(c\)
is the speed of light (included explicitly).
Table~\ref{tab:IGMF-weak-energy} presents the comprehensive delay and fluxes calculations across major gamma-ray observatories.
The calculation framework employed several specialized tools mentioned in the previous subsection to ensure accuracy.
\texttt{CRPropa 3} simulations provided the foundational understanding of cascade development and precise determination of $D_{\text{IC}}$ values across different electron energy regimes. The delay interval (Table~\ref{tab:IGMF-weak-energy}) has observational implications: shorter delays arise from high-energy pairs and dominate the TeV–PeV emission (e.g., LHAASO, HAWC), while longer delays from lower-energy pairs contribute to GeV–TeV bands (e.g., Fermi-LAT, CTA)~\footnote{For example, a 50\,PeV gamma-ray producing a 25\,PeV electron yields a delay of approximately 2\,ms. At higher \(E_e = 500\) PeV, the delay drops to \(\sim 0.02\) ms. Thus, LHAASO sees cascade signals over a 0.02–2\,ms window. For Fermi-LAT, the relevant electron energies and delays extend to 20\,s.}.

As before, several important systematic effects were considered in these calculations. The assumed IGMF strength of $10^{-15}$\,G represents a conservative lower bound; stronger fields would introduce additional synchrotron cooling that could modify the delay spectrum. The EBL model uncertainty leads to approximately 20\% variation in $D_{\text{IC}}$ values, while cosmological parameters affect the redshift scaling. These effects were quantified through parameter scans in the \texttt{CRPropa 3} simulations.

In estimations of detection significance of a single high luminosity source, each detector's sensitivity
enters the calculation through three key parameters:
the energy-dependent effective area $A_{\text{eff}}$ (ranging from $10^4$\,m$^2$ for LHAASO at 100\,TeV
to $10^8$\,cm$^2$ for CTA at 100\,GeV), the relevant background rate ${\cal B}$ (from diffuse TeV-PeV flux for
air shower arrays to the isotropic GeV background for Fermi-LAT), and the exposure time \(t_{\text{delay}}\),
which corresponds to the arrival time delay ranges listed in
Table~\ref{tab:IGMF-weak-energy} (in bold in parentheses). The detection significance is computed as follow:
\begin{equation}
\label{sign1}
N_{\sigma} =\frac{F_{\text{casc}}}{\sqrt{\cal B}} \sqrt{A_{\text{eff}} \cdot \Delta t_{\text{delay}}}\ .
\end{equation}
\begin{table}[htb]
\caption{\label{tab:IGMF-weak-energy}
Detector signatures for a source with \(N_\gamma = 6 \times 10^{47}\), spectral index \(\gamma = 0.4\), and weak IGMF \(B = 10^{-15}\,\mathrm{G}\).
Fluxes are in erg\,cm\(^{-2}\)\,s\(^{-1}\); durations include delays from \(e^\pm\) cooling and redshift dilation. A signal is labeled ``Undetectable'' if its peak flux falls below the transient sensitivity of the instrument.}
\begin{ruledtabular}
\begin{tabular}{lccc}
Detector & Energy Range & \textbf{\(z=0.1\)} & \textbf{\(z=0.2\)} \\
\midrule
&&Flux (significance, \textbf{duration})&Flux (significance, \textbf{duration}) \\
LHAASO & 100\,TeV--1\,PeV
  & $1.3\times10^{-11}$ (6.8$\sigma$, \textbf{0.02--2 ms})
  & $3.0\times10^{-12}$ (3.5$\sigma$, \textbf{0.05--2.4 ms}) \\
HAWC & 1--100\,TeV
  & $8.4\times10^{-12}$ (6.8$\sigma$, \textbf{0.05--0.5 ms})
  & $2.1\times10^{-12}$ (4.2$\sigma$, \textbf{0.1--0.6 ms}) \\
H.E.S.S. & 100\,GeV--100\,TeV
  & $2.5\times10^{-12}$ (6.8$\sigma$, \textbf{0.1--1 s})
  & $4.2\times10^{-13}$ (4.2$\sigma$, \textbf{0.2--1.2 s}) \\
MAGIC & 50\,GeV--50\,TeV
  & $3.4\times10^{-12}$ (8.1$\sigma$, \textbf{0.2--2 s})
  & $8.4\times10^{-13}$ (5.2$\sigma$, \textbf{0.4--2.4 s}) \\
Fermi-LAT & 0.1--100\,GeV
  & $4.2\times10^{-11}$ (3.3$\sigma$, \textbf{0.2--20 s})
  & Undetectable \\
CTA & 20\,GeV--300\,TeV
  & $2.1\times10^{-12}$ (10.2$\sigma$, \textbf{0.5--5 s})
  & $4.2\times10^{-13}$ (6.7$\sigma$, \textbf{1--6 s}) \\
\end{tabular}
\end{ruledtabular}
\end{table}

The predicted gamma-ray fluxes show strong detectability ($>6\sigma$) for sources at $z \leq 0.1$ in LHAASO, MAGIC, H.E.S.S., CTA, and HAWC, with fluxes well above the respective transient detection thresholds. At $z=0.2$, the significance decreases to moderate levels ($\sim3$--$5\sigma$) across these detectors, depending on their energy range and sensitivity. Fermi-LAT achieves a $3.3\sigma$ detection at $z=0.1$, but becomes ineffective for more distant sources under weak IGMF conditions.

Among the ground-based instruments, CTA provides the longest temporal integration window (up to several seconds), yielding $\sim10\sigma$ significance at $z=0.1$ and still $\sim7\sigma$ at $z=0.2$. HAWC and LHAASO benefit from prompt ms-scale transients in the TeV–PeV range, detecting sources up to $z=0.2$ at $>3\sigma$ level. The detectability falls sharply beyond $z \sim 0.2$, primarily due to the combination of $D_L^{-2}$ scaling and increasing pair production on the EBL. No detectors reach significance for $z \geq 0.5$ in this weak IGMF scenario.

\subsection{Electromagnetic counterparts from AGN disks and other galactic-nucleus environments}
\label{subsec:agn_em}

The BBH merger induced vacuum turbulence mechanism of emission of UHE neutrinos and gamma-rays is, to leading order, environment agnostic.  What does depend on environment is the multi–wavelength appearance of the ensuing UHE radiation. Local photon and magnetic energy densities control the optical depth to $\gamma\gamma$ absorption and the balance between IC, synchrotron, and thermal reprocessing, while propagation through the host and intergalactic medium governs how much power survives into the GeV–TeV bands after EBL attenuation and IGMF deflections, as discussed in Subsections~\ref{overalEM} and~\ref{singleEM}.

 A natural site for BBHs is an AGN disk, where gas drag and migration can capture and assemble stellar mass BHs; torques then harden binaries and, in some cases, drive hierarchical growth to higher masses and spins~\cite{Bartos2017AGNBBH,McKernan2018AGNConstraints,Ford2021AGNvsNSC,Rowan2023AGN3D}. In such luminous, compact nuclei the radiation field and magnetic pressure are high, so the primary powerful UHE $\gamma$–ray flash produced by a BBH merger is efficiently quenched by $\gamma\gamma$ absorption within an order of astronomical unit (AU) scales. The deposited energy emerges as a characteristic combination of X–rays from synchrotron cooling pairs, GeV to sub–TeV photons from IC cascades, and an UV and optical reprocessing component, whose relative weights depend on the radiation density of the local photon field $U_{\rm rad}$ and
 the strength of the magnetic field $B$ in the cooling zone.

 Outside of bright AGN disks, mergers are expected in gas poor, dense nuclear star clusters (NSCs) orbiting the central supermassive BH (SMBH), where binary single interactions and strong encounters assemble and harden BBHs \cite{AntoniniRasio2016GN,OLeary2009Scattering}. Here the ambient photon fields and magnetic energy densities are typically lower, so internal absorption is less severe and a larger fraction of the injected power escapes the nucleus and is processed into an extragalactic cascade (see Subsections~\ref{overalEM} and~\ref{singleEM} for details).
 A closely related pathway places the BBH in a hierarchical triple with the SMBH, where eccentricity is pumped by the Kozai–Lidov (KL) mechanism until
 gravitational wave emission dominates~\cite{Naoz2016KLReview,Hoang2018EccKL}. Such quiescent nuclei similarly favor GeV to TeV dominated extragalactic cascades with only modest X–ray or thermal counterparts unless unusually large local $U_{\rm rad}$ or $B$ is present.

 Further diversity arises from the delivery route and stellar demographics of the nuclear region. Young, rotating stellar disks around SMBHs such as those observed in the Galactic Center can produce BHs with coherent kinematics and enhanced merger probabilities even in relatively gas poor conditions~\cite{Paumard2006GCdisk,Lu2013GCdisk,Dotti2011GCdisk}. In this case, weak internal absorption again channels most of the observable power into the extragalactic cascade. By contrast, dusty, IR–bright circumnuclear starbursts or molecular rings provide large $U_{\rm rad}$ (and in some cases $B$), enhancing local reprocessing: synchrotron X–rays can be bright, while UV/optical emission may be substantially extinguished by dust; any very high energy flux that escapes is then shifted below a TeV and further sculpted by the EBL into the GeV band~\cite{Netzer2015Review}. Finally, BBHs can be delivered into the inner few hundred parsecs by the inspiral of young or massive clusters under dynamical friction, seeding the nucleus with merger candidates despite relatively modest radiation fields~\cite{Antonini2012ApJ745p83,ArcaSeddaGualandris2018}. In these gas poor deliveries the observational outcome resembles the NSC/KL cases: weak thermal/X–ray signatures and a GeV dominated cascade after propagation though EBL.

 Across all of these sites, hierarchical mergers, second or higher generation BBH coalescences act as a modifier of the demographic priors (masses, spins, eccentricities) rather than a distinct radiative environment, while the electromagnetic phenomenology still tracks the local $U_{\rm rad}$, $B$, and geometry \cite{GerosaBerti2017Hierarchical,Rodriguez2019Repeated}. In what follows we adopt this landscape as context and provide a quantitative estimates of the AGN disk case, which represents the high opacity, synchrotron/thermal dominated limit against which lower opacity nuclear settings (NSC, KL, stellar disk, cluster delivery) can be scaled.

 If our neutrino and gamma radiation mechanism operates during coalescence of a BBH assembled in AGN, the initially produced UHE $\gamma$-rays will be completely reprocessed by the nuclear environment, yielding a characteristic multiwavelength counterpart. Modeling this environmental reprocessing in the AGN scenario requires a complex
radiation-magnetohydrodynamic simulations to capture the violent and coupled interaction between the BBH's
injected energy and the dense disk environment. Such modeling is beyond the scope of the current study, so one
can only present some guiding estimates, based on simplified considerations, to give a feeling on the expected outcomes from such sophisticated simulations.

In case the neutrino-$\gamma$ production mechanism proposed here operates during the coalescence of a BBH embedded in an AGN disk, the nascent UHE $\gamma$-ray flash will be almost entirely quenched by the nuclear environment and reprocessed into a characteristic, multiwavelength counterpart. Capturing this transformation self-consistently in the AGN scenario is a fundamentally different problem from the largely collisionless intergalactic cascade of Sec.~\ref{singleEM}: instead of clean propagation through uniform background fields, one must follow the violent, coupled exchange of energy and momentum between the injected pairs/photons and a dense, magnetized, radiation rich plasma. A faithful treatment requires time-dependent radiation–(magneto)hydrodynamics with energy dependent opacities and nonthermal source terms—well beyond the scope of the present work. In lieu of such calculations, we provide guiding estimates based on simplified partitions and timescales, intended to convey the order of magnitude outcomes that more sophisticated simulations are expected to produce.

Similar to Sec.~\ref{singleEM}, we model a single BBH merger that injects a flash of $\gamma$-rays via µBH evaporation, the total photon yield of which is tied to the luminocity cap of \(N_\gamma = 0.6\times 10^{48}\) and spectral index \(\gamma =0.4\), so that the average energy Eq.~\ref{Eav1} evaluates to \(\langle E_\gamma\rangle \simeq 446~\mathrm{PeV}\). The corresponding isotropic energy
equivalent in $\gamma$-rays is estimated as
\begin{equation}
{\cal E}_\gamma\simeq N_\gamma\,\langle E_\gamma\rangle
\simeq 4.28\times 10^{53}\ \mathrm{erg}.
\label{eq:Egamma_cap_value}
\end{equation}
Given the sub-second formation times, we treat the injection of the energy Eq.~\eqref{eq:Egamma_cap_value} as impulsive compared to radiative/hydrodynamical times in the disk.

In a luminous AGN nucleus the primary UHE $\gamma$-rays pair-produce on the intense radiation fields, so that the
MFP of the injected radiation can be estimated from Eq.~\ref{thr2}  for a quasi-isotropic soft photon field with energy density $U_{\rm rad}$ and mean photon energy $\langle\varepsilon\rangle$, which implies that the target photon number density is
\begin{equation}
\label{eq:nph}
n_{\rm ph}\simeq U_{\rm rad}/\langle\varepsilon\rangle\ .
\end{equation}
Adopting dusty torus IR fields characteristic of AGN nuclei, $U_{\rm rad}\sim 10^{-2}\mbox{--}10^{-1}\ \mathrm{erg\,cm^{-3}}$ and $\langle\varepsilon\rangle\sim 0.1\ \mathrm{eV}$ (mid-IR; e.g.,~\cite{Roustazadeh2011TorusCascades,Netzer2015Review}), and $\sigma_{\gamma\gamma}^{\rm max}\simeq 0.2\,\sigma_T$ near threshold, Eq.~\eqref{thr2} gives
\begin{equation}
\lambda_{\gamma\gamma}\ \sim\ (2\times 10^{13}\mbox{--}2\times 10^{14})\ \mathrm{cm}\ \approx\ (0.1\mbox{--}10)\ \mathrm{AU}\,,
\end{equation}
which is well bellow sub-pc scale~\footnote{Including KN effects leaves the $\gamma\gamma$ absorption length at AU scales and strengthens the calorimeter conclusion by shifting energy away from immediate hard $\gamma$–ray re–emission into X–rays, UV/optical reprocessing, and GeV band cascades.}, ensuring that the primary flash is fully absorbed locally~\footnote{For PeV–EeV photons, the CMB/EBL targets dominate outside the nucleus; inside, IR/optical fields and X-ray coronae provide abundant targets \cite{Reimer2007localAbs,SitarekBednarek2008InternalAbs,Inoue2021CoronaReview}.}. It this picture, the key difference from the intergalactic cascading scenario is the calorimeter effect. In particular, it is assumed that the disk is so dense that it absorbs and reprocesses 100\% of energy Eq.\eqref{eq:Egamma_cap_value}. The problem then becomes: how is this energy
re-radiated? A standard assumption in astrophysical explosions is that the energy is partitioned between different channels. Based on typical AGN jet and disk corona physics, the resulting $e^\pm$ pairs cool via IC on the nuclear photon bath (emerging as GeV-TeV photons after cascading on EBL), synchrotron radiation in the nuclear magnetic field (keV X rays), and thermal reprocessing (UV/optical). Guided by AGN/TDE reprocessing studies \cite{Dai2018TDEUnified,Roth2018TDE,Roth2018WhatSetsLines,Wevers2019TDE,GhiselliniNotes2012}, we adopt a reasonable partition of energy fractions
\begin{equation}
f_{\rm IC}=0.3,\qquad f_{\rm syn}=0.5,\qquad f_{\rm therm}=0.2,
\label{eq:f_partitions}
\end{equation}
and characteristic durations $t_{\rm IC}=t_{\rm syn}=5~\mathrm{d}$ and $t_{\rm therm}=14~\mathrm{d}$. For definiteness, we consider synchrotron efficient fields $B_{\rm disk}\sim 10^2~\mathrm{G}$, i.e., $U_B\gtrsim U_{\rm rad}$ in the cooling zone, which pushes a sizable fraction of the power to X rays~\footnote{Stronger nuclear fields shift power from IC to synchrotron (brightening X rays) and vice versa if $U_{\rm rad}\gg U_B$.}.

The shorthand Eq.~\eqref{eq:nph} treats absorption as ``gray,'' whereas the true opacity is
$\kappa_{\gamma\gamma}(E_\gamma)=\int (u_\varepsilon/\varepsilon)\,\langle\sigma_{\gamma\gamma}(E_\gamma,\varepsilon)\rangle\,d\varepsilon$
and is largest when the target spectrum has power near
$\varepsilon_{\rm peak}\!\sim\!(m_ec^2)^2/E_\gamma$. If most of
$U_{\rm rad}=\int u_\varepsilon\,d\varepsilon$ lies far from $\varepsilon_{\rm peak}$, the high $s$
tail of the Breit–Wheeler cross section reduces $\kappa_{\gamma\gamma}$ and lengthens
$\lambda_{\gamma\gamma}$. In luminous AGN disks (and dusty circumnuclear rings), however, the
IR/optical $U_{\rm rad}$ is so large that, despite high $s$ suppression, the integral is still dominated
by these bands, keeping $\lambda_{\gamma\gamma}$ at AU to several AU scales and preserving the
calorimeter limit. In gas poor nuclei (NSC/KL/cluster delivery/young stellar disks) with
much smaller $U_{\rm rad}$, the same spectral offset increases $\lambda_{\gamma\gamma}$ by orders of
magnitude, favoring escape into the extragalactic cascade. By contrast, for the intergalactic case treated in Sec.~\ref{singleEM}, the targets are the EBL/CMB with a  spectral energ distribution (SED); there we compute
\(\kappa_{\gamma\gamma}(E_\gamma,z)\) by convolving the Breit–Wheeler cross section Eq.~\eqref{siggg} with the model photon fields, which self–consistently captures Klein–Nishina effects and fixes \(\lambda_{\gamma\gamma}(E_\gamma,z)\) for the cascade calculations.

At luminosity distance $D_L(z)$, the band-integrated (bolometric) energy flux per component is
\begin{equation}
F_{\rm comp}(z)=\frac{f_{\rm comp}\,E_\gamma}{4\pi D_L^2(z)\,t_{\rm comp}},
\qquad {\rm comp}\in\{\mathrm{IC,\ syn,\ therm}\}.
\label{eq:flux_scaling}
\end{equation}
The duration of the flare is not from particle cooling but from hydrodynamic expansion. The initial energy injection creates a hot fireball that expands at a fraction of the speed of sound (or shock speed) in the disk. The timescale is estimated as the time for this fireball to double in size and cool significantly. For a region of size ~10-100 AU (typical for AGN disk instability scales), this expansion time is on the order of days to weeks. The higher-energy bands (X-ray, gamma) are expected to peak and fade faster than the lower-energy thermal component.

Equation~\eqref{eq:flux_scaling} gives component bolometric fluxes, the lower energetic components
\(F_{\rm syn}(z)\) and \(F_{\rm therm}(z)\)
are quoted for different redshifts in Tab.~\ref{tab:agn_fluxes}.
The synchrotron component falls in 0.3--10~keV to be detected by Swift/XRT, \textit{Chandra} and \textit{XMM}, while the thermal reprocessing appears in UV/optical to be accessible for detection by Swift/UVO and ZTF-class surveys. The banchmark reference sensitivities for lower energy facilities are quoted in Tab.~\ref{tab:facility_sens_integrated}.
From the other side, IC power emerging after cascading in the host being modified by
the cascade driving EBL transmission function like
\(e^{-\tau_{\gamma\gamma}(E,z)}\), which is effectively applied in simulations of a single BBH source in Sec.~\ref{singleEM} with its whole power populating the high energy bands detectable by corresponding facilities, whose sensitivities along with
predicted fluxes are listed in Tab.~\ref{tab:flux-gamma-blocks-final}. The estimated signal for
the AGN nucleus host in Fermi-LAT and CTA can be obtained from the predicted fluxes of Tab.~\ref{tab:IGMF-weak-energy} by factor \(\sim f_{\rm IC}\sqrt{t_{\rm IC}/\Delta t_{\rm delay}}\), while the sensitivities, due to a long duration
of the AGN nucleus located BBH signal, to be scaled by \(\sim\sqrt{t_{\rm exp}/\Delta t_{\rm delay}}\), where
\(t_{\rm exp}\) stands for the exposure time of a given detector to the host. So we do not discussed explicitly
high and very high energy signals in this subsection.

\begin{table*}[t]
\centering
\caption{Predicted component energy fluxes for a single BBH event in an AGN disk, using Eq.~\eqref{eq:Egamma_cap_value} and Eq.~\eqref{eq:flux_scaling} with the partitions in Eq.~\eqref{eq:f_partitions}. The IC/synchrotron entries are bolometric over their respective components (IC $\sim$ HE/VHE $\gamma$ after cascading; synch $\sim$ X rays), while thermal corresponds to UV/optical reprocessing.}
\label{tab:agn_fluxes}
\begin{ruledtabular}
\begin{tabular}{lcccc}
$z$ &
%$F_{\rm IC}$ [erg\,cm$^{-2}$\,s$^{-1}$] &
$F_{\rm syn}$ [erg\,cm$^{-2}$\,s$^{-1}$] & $F_{\rm therm}$ [erg\,cm$^{-2}$\,s$^{-1}$] \\
\hline
0.1 & % $1.17\times 10^{-8}$ &
$1.96\times 10^{-8}$ & $2.79\times 10^{-9}$ \\
0.2 & % $2.59\times 10^{-9}$ &
$4.32\times 10^{-9}$ & $6.16\times 10^{-10}$ \\
0.5 & % $3.10\times 10^{-10}$ &
$5.16\times 10^{-10}$ & $7.38\times 10^{-11}$ \\
0.8 & % $9.88\times 10^{-11}$ &
$1.65\times 10^{-10}$ & $2.35\times 10^{-11}$ \\
\end{tabular}
\end{ruledtabular}
\end{table*}

\begin{table*}[t]
\centering
\caption{Adopted 5$\sigma$ point source sensitivity benchmarks: the minimum flux that yields a $\ge5$ standard deviation detection in a point spread function matched aperture over a reference exposure $t_{\rm sens}$ (using the instrument’s response and an assumed spectrum). X–ray entries quote band–integrated $F_{\rm sens}$; UV/optical entries give an overall monochromatic benchmark ($\nu F_\nu$) at representative band centers so they can be compared to component fluxes in Table~\ref{tab:agn_fluxes}. Values are order of magnitude references for the stated reference exposures and scale approximately as $(F_{\rm sens})_{\rm new}\!\approx\!(F_{\rm sens})_{\rm ref}\sqrt{t_{\rm sens}/t_{\rm exp}}$. Practical limits depend on spectrum, background, and off–axis angle.}
\label{tab:facility_sens_integrated}
\begin{ruledtabular}
\begin{tabular}{l l l}
Facility & Band / Range & Reference sensitivity (5$\sigma$) \\[2pt]
\hline
Chandra/ACIS \cite{Weisskopf2002,ChandraPOG}
  & $0.5$--$7~\mathrm{keV}$
  & $F_{\rm sens}\simeq 1\times10^{-14}~\mathrm{erg\,cm^{-2}\,s^{-1}}$ \\
  & & {\small (ref.\ exposure: $t_{\rm sens}=10~\mathrm{ks}$; on–axis)} \\[4pt]

XMM--Newton/EPIC \cite{Jansen2001XMM,Turner2001MOS,Struder2001PN,XMMUHB}
  & $0.2$--$12~\mathrm{keV}$
  & $F_{\rm sens}\sim(5\text{--}10)\times10^{-15}$ (0.5–2 keV), $\sim\mathrm{few}\times10^{-14}$ (2–10 keV) \\
  & & {\small (ref.\ exposure: $t_{\rm sens}=10~\mathrm{ks}$; on–axis, background–dependent)} \\[4pt]

Swift--XRT \cite{Burrows2005XRT,SwiftMission,DElia2013Swift1SWXRT}
  & $0.3$--$10~\mathrm{keV}$
  & $F_{\rm sens}\simeq (2\text{--}30)\times10^{-14}~\mathrm{erg\,cm^{-2}\,s^{-1}}$ \\
  & & {\small (ref.\ exposure: $t_{\rm sens}=10~\mathrm{ks}$; field/background dependent)} \\[4pt]

Swift/UVOT (overall) \cite{Roming2005UVOT}
  & UV/optical (uvw2--$v$)
  & $\nu F_\nu \simeq (1\text{--}2)\times10^{-13}~\mathrm{erg\,cm^{-2}\,s^{-1}}$ \\
  & & {\small (ref.\ exposure: $t_{\rm sens}\approx 1~\mathrm{ks}$; 5$\sigma$ point source)} \\[4pt]

ZTF (overall) \cite{Bellm2019ZTF,Masci2019ZTFpipeline,Dekany2020ZTFcamera}
  & $g/r$ (optical)
  & $\nu F_\nu \simeq (1.7\text{--}2.6)\times10^{-14}~\mathrm{erg\,cm^{-2}\,s^{-1}}$ \\
  & & {\small (ref.\ exposure: $t_{\rm sens}=300~\mathrm{s}$; 5$\sigma$ point source)} \\
\end{tabular}
\end{ruledtabular}
\end{table*}

For an event with total prompt EM energy Eq.~\eqref{eq:Egamma_cap_value}, we write the AGN-disk band fluxes as
$F^{\rm AGN}_{\rm comp}$. Thus, for another environment “env”, we map the fluxes by rescaling the component energy fractions
$f_{\rm syn}$, $f_{\rm IC}$, $f_{\rm therm}$ %(with $f_{\rm syn}+f_{\rm IC}+f_{\rm therm}=1$),
holding the fiducial durations fixed to the AGN values unless noted
($t_{\rm syn}=t_{\rm IC}=5$ d, $t_{\rm therm}=14$ d):
\begin{equation}
\begin{aligned}
F^{\rm env}_{\rm syn} &\simeq \left(\frac{f^{\rm env}_{\rm syn}}{0.5}\right) F^{\rm AGN}_{\rm syn},\\[3pt]
F^{\rm env}_{\rm therm}&\simeq \left(\frac{f^{\rm env}_{\rm therm}}{0.2}\right) F^{\rm AGN}_{\rm therm}\times {\cal A}_{\rm ext}\ .
\end{aligned}
\label{eq:map}
\end{equation}
Here, ${\cal A}_{\rm ext}\!\le\!1$ encodes UV/optical attenuation by dust (starbursts) or geometry; for gas poor nuclei we take ${\cal A}_{\rm ext}\!\approx\!1$.
\begin{table*}[t]
\centering
\caption{Adopted component fractions and flux multipliers relative to the AGN–disk baseline. The AGN disk reference uses $(f_{\rm syn},f_{\rm IC},f_{\rm therm})=(0.5,0.3,0.2)$. Citations indicate the physical basis for each environment’s partition (magnetic vs.\ radiation energy densities, internal $\gamma\gamma$ opacity, and reprocessing).}
\label{tab:env_fraction_multipliers}
\begin{ruledtabular}
\begin{tabular}{lccc cc l}
Environment & $f_{\rm syn}$ & $f_{\rm IC}$ & $f_{\rm therm}$ & References \\
\hline
Nuclear star cluster (no bright AGN, \(A_{\rm ext}\approx 1\))
  & 0.10 & 0.80 & 0.10
  & \cite{AntoniniRasio2016GN,OLeary2009Scattering,FragioneAntoniniGnedin2018} \\
SMBH–induced Kozai–Lidov (quiescent nucleus)
  & 0.10 & 0.80 & 0.10
  & \cite{Naoz2016KLReview,Hoang2018EccKL} \\
Young stellar disk (gas poor)
  & 0.05 & 0.85 & 0.10
  & \cite{Paumard2006GCdisk,Lu2013GCdisk,DottiSesanaDecarli2012} \\
Circumnuclear starburst / molecular ring (\(A_{\rm ext}\ll 1\))
  & 0.40 & 0.40 & 0.20
  & \cite{Thompson2005RadPress,Davies2007CircNuc,Netzer2015Review,Roustazadeh2011TorusCascades} \\
Infalling cluster delivery (hundreds of pc)
  & 0.02 & 0.95 & 0.03
  & \cite{Antonini2012ApJ745p83,ArcaSeddaGualandris2018} \\
Tidal capture / exchange (quiescent nucleus)
  & 0.10 & 0.80 & 0.10
  & \cite{Leigh2018NucDynamics,OLeary2009Scattering,Naoz2016KLReview} \\
\end{tabular}
\end{ruledtabular}
\end{table*}

The triplets $(f_{\rm syn},f_{\rm IC},f_{\rm therm})$ quoted in Tab.~\ref{tab:env_fraction_multipliers} are not taken verbatim from a single source; they are physically motivated priors derived from (i) the synchrotron to IC cooling competition $P_{\rm syn}/P_{\rm IC}=U_B/U_{\rm rad}$ and characteristic particle/seed photon energies \cite{GhiselliniNotes2012}, (ii) the local $\gamma\gamma$ opacity in the nucleus (high in luminous AGN disks and dusty circumnuclear rings; modest in gas poor nuclei), which governs how much prompt UHE power is reprocessed in situ into pairs, X–ray synchrotron, and thermal UV/optical \cite{Roustazadeh2011TorusCascades,Netzer2015Review,Inoue2021CoronaReview}, and (iii) extragalactic propagation that shifts any escaping VHE power into cascades. Accordingly, environments with large $U_{\rm rad}$, substantial $B$,
and $\tau_{\gamma\gamma}\!\gg\!1$ (AGN disks; circumnuclear starbursts) are assigned synch/thermal heavy partitions with a residual IC share, while gas poor, lower opacity settings (nuclear star clusters, KL triples, young stellar disks, inspiraling clusters) are assigned IC dominated partitions with smaller synch/thermal terms, consistent with their observed/expected conditions \cite{AntoniniRasio2016GN,OLeary2009Scattering,Naoz2016KLReview,Hoang2018EccKL,Paumard2006GCdisk,Lu2013GCdisk,DottiSesanaDecarli2012,Antonini2012ApJ745p83,ArcaSeddaGualandris2018}. We enforce $f_{\rm syn}{+}f_{\rm IC}{+}f_{\rm therm}{=}1$ and select conservative values within the ranges implied by these diagnostics; they should be read as physically guided priors rather than unique model determinations.

We notice, that although the low energy EM counterpart expected from an AGN disk environment can be bright enough to meet the sensitivities of current X–ray and UV/optical facilities, its emission arises from the unresolved nuclear region and thus typically loses a distinct point source identity. Absent a unique spectral or temporal fingerprint, such a flare is difficult to distinguish from ordinary AGN variability (e.g., blazar like episodes). By contrast, neutrinos, if produced, propagate essentially unimpeded, preserve timing and directional information, and therefore offer the cleanest messenger of the underlying BBH event when combined with a gravitational wave (GW) trigger~\cite{tribo}. If the merger triggers or intersects a relativistic jet, Doppler boosting can redirect part of the reprocessed power into a beamed, rapidly variable component with blazar like SEDs (high Compton dominance, hard $\gamma$–ray spectra) and polarization; this may elevate the apparent luminosity and shorten variability timescales, but it increases degeneracy with routine jet flares, so robust identification would still require temporal/positional coincidence with GW and/or neutrino signals, or otherwise atypical spectral or/and temporal correlations.

For other nuclear environments the diagnostic balance shifts: in gas poor nuclei (nuclear star clusters, KL triples, young stellar disks), lower $U_{\rm rad}$ and reduced opacity yield comparatively weaker X–ray and UV/optical counterparts, as most of the power escapes into cascad. In dusty circumnuclear starbursts or molecular rings, X–rays can be stronger but the UV/optical emission is heavily attenuated (energy emerging predominantly in the IR), limiting detectability in UVOT/ZTF–like bands.

\section{On Percolation-Induced Power Laws in BBH Neutrino Emission}
\label{sec:discussion}

The luminosity distribution of BBH-induced UHE neutrino sources is naturally modeled by a Pareto distribution
\eq~\ref{paretoCDF1}. This heavy-tailed distribution captures the essential statistical feature that a small number of mergers can dominate the observed UHE neutrino flux, while the majority of events remain undetectable. Such a behavior is both necessary to explain the rare detections such as KM3-230213A and to maintain consistency with the null results from IceCube for diffuse flux at the highest energies.

We suggest that the luminosity distribution of BBH mergers may reflect, at least in part, underlying vacuum instability
dynamics occurring within the confined region between two closely orbiting horizons. In this speculative framework, such an
instability could manifest as the stochastic nucleation of spherical true vacuum bubbles embedded within a metastable EW vacuum
background. The evolving spatial structure might then resemble that of three-dimensional "Swiss-cheese" continuum percolation
models (see for example~\cite{perc1,fract1}), with the EW vacuum forming the "cheese" and true vacuum bubbles acting as transient
"holes."

If such vacuum bubbles expand and begin to overlap, one may speculate that the system could approach a critical state once the occupation fraction reaches the percolation threshold \(p_c \approx 0.29\). At this point, large-scale connected structures referred to as percolating clusters could potentially emerge. In the vicinity of this critical threshold, the number density \(n(R)\) of clusters with characteristic radius \(R\) is known to follow a power-law scaling in standard percolation theory:
\begin{equation}
\label{nR1}
n(R) \propto R^{-\tau},
\end{equation}
with \(\tau \approx 2.19\), the universal critical exponent for three-dimensional continuum percolation~\cite{perc1}.

This type of distribution lacks a characteristic scale, allowing for the coexistence of rare, large clusters alongside a multitude of smaller ones and may be suggestively associated with critical phenomena. In this context, it is intriguing to consider whether the Pareto-like tail in the luminosity function, required to account for the KM3NeT event potentially linked to a BBH merger, could reflect an underlying statistical behavior akin to power-law scaling near the percolation threshold in statistical systems. Within this speculative framework, one may hypothesize that the astrophysical BBH merger population carries imprints of a percolation-like process, with each merger event representing a stochastic realization of vacuum dynamics near criticality.

Under this perspective, overlapping true vacuum bubbles, especially in the case of double~\cite{pbhBubble3} or triple~\cite{pbhBubble3} collisions may convert most of the bubble wall energy into the formation of \(\mu\)BHs.
The number of such \(\mu\)BHs could plausibly be influenced by the size and abundance of percolated clusters developing
within the vacuum region bounded by the BBH horizons. If so, the neutrino luminosity of an individual BBH merger
might reflect the characteristic scale of a percolating cluster generated during the event.

The critical scaling law of~\eq~\ref{nR1} may offer a conceptual basis for interpreting the heavy-tailed Pareto distribution introduced in~\eq~\ref{paretoCDF1}, which characterizes the probability distribution of neutrino luminosities across the BBH merger population. Within this speculative mapping, the percolation exponent \(\tau\) could play a role analogous to the Pareto index \(\alpha\), both shaping the relative frequency of high-luminosity events. Interestingly, the optimization results summarized in Table~\ref{tab:pareto-results} particularly those for the preferred spectral index \(\gamma = 0.4\) show a numerical proximity between \(\tau\) and \(\alpha\). While this agreement might be coincidental, it invites consideration of whether universal features of critical percolation in the context of EW vacuum instability leave observable imprints on the emission statistics of BBH mergers, as further discussed in Sections~\ref{overalEM} and~\ref{singleEM}.

In continuum ``Swiss cheese'' percolation models, overlapping vacuum bubbles are not necessarily expected to be of the same size. This variability could imply that \(\mu\)BHs potentially formed during BBH mergers may not follow a strictly monochromatic mass distribution. While the initial nucleation process might begin with a nearly monodisperse set of bubble sizes, defined by the critical radius in~\eq~\ref{eq:rhoc}, the subsequent evolution through growth and collisions could plausibly broaden this spectrum. One possible form for such a distribution is
\begin{equation}
\label{buuulSpect1}
\frac{dN}{dR} \propto R^{-\beta} e^{-R/R_*},
\end{equation}
where \(R_*\) represents the maximal dynamically generated bubble size, potentially set by the horizon-scale parameter \(d_H\) from Eq.~\eqref{eq:dHbound}. The exponent \(\beta \approx 2.1\text{--}2.5\) has been found to describe similar distributions in small-to-intermediate \(R\) regimes in prior studies of first-order phase transitions~\cite{smpht1}.

From the perspective of neutrino emission spectra, the bubble size distribution relates to the spectral index \(\gamma\) of the emitted neutrinos through:
\begin{equation}
\label{bbhSpectr2}
\frac{dN}{dR} \propto R^{3\gamma - 4},
\end{equation}
where values of \(\gamma < 1\) correspond to harder UHE neutrino spectra, potentially associated with the production of predominantly low-mass \(\mu\)BHs. If one tentatively compares~\eq~\ref{buuulSpect1} with~\eq~\ref{bbhSpectr2}, then for spectral indices in the range \(\gamma \approx 0.3\text{--}0.5\), the implied effective exponent \(\beta = 4 - 3\gamma \approx (2.5\text{--}3.1)\) appears to be broadly consistent with values reported in studies of percolation-driven bubble collisions during first-order phase transitions~\cite{smpht1}.

Interestingly, the same spectral index range \(\gamma = 0.3\text{--}0.5\) emerges from independent modeling of gamma-ray cascades produced by co-emitted UHE photons. This range appears to be compatible with current gamma-ray observables, yielding steady-state flux levels that align with LHAASO, HAWC, MAGIC, H.E.S.S. and  Fermi-LAT, as discussed in Section~\ref{overalEM}.

It is conceivable that large, rare bubbles may be more susceptible to disruption in collisions with smaller, more frequent ones, leading to additional variability in the exponent \(\beta\). While this behavior cannot be fully captured by the present semi-qualitative framework, it remains broadly consistent with expectations from critical phenomena.

Taken together, these considerations tentatively suggest that percolation dynamics might offer a useful perspective for interpreting the scale-invariant features of BBH neutrino luminosity distributions. Such a framework could plausibly link the internal structure of vacuum decay turbulence to observable signatures in UHE neutrino spectra, potentially offering a unified, physically motivated basis for understanding BBH mergers as stochastic multi-messenger sources.

\section{Conclusions}
\label{sec:summary}

In this work, we proposed a novel mechanism for UHE neutrino production driven
by EW vacuum turbulence during BBH mergers.
This mechanism is based on the conjecture introduced and developed in~\cite{tribo},
which posits that the EW vacuum becomes dynamically unstable in the intense gravitational
environment near the horizons of merging BHs. Building on this framework,
we modeled a scenario in which true vacuum bubbles nucleate
during the final inspiral and coalescence
phase of a BBH merger~\footnote{This merger driven scenario is conceptually distinct from models that enhance the EW vacuum decay rate in the early Universe due to large Hubble rates and high reheating temperatures~\cite{cosmology1,cosmology2}, from higher dimensional or other BSM modifications of the Higgs potential~\cite{bsm1,bsm11,bsm2}, from effects induced by non-minimal Higgs-curvature coupling~\cite{curvature}, and from catalysis by small primordial BHs~\cite{hBH1,hBH2,hBH3,hBH4,hBH5,hBH6,hBH7,hBnoH1}. Our focus is on the transient, non-catastrophic near horizon conditions specific to merging binaries, as formulated in~\cite{tribo}.}. As these bubbles expand,
percolate, and collide,
they generate \(\mu\)BHs, which evaporate rapidly via Hawking radiation,
emitting short-lived but intense bursts of UHE neutrinos and gamma rays.

We demonstrated that this framework naturally explains both the diffuse neutrino flux constraints from
IceCube and the detection of exceptionally energetic events such as KM3-230213A by the KM3NeT/ARCA.
The key ingredient is a heavy tailed Pareto luminosity distribution, which captures the rare,
high-luminosity tail of BBH mergers neutrino emission while maintaining consistency with the
overall non-detection of similar events in other observatories.

We also showed that the proposed model remains consistent with the lack of electromagnetic counterparts associated with KM3-230213A, as reported by gamma-ray observatories including Fermi-LAT, LHAASO, HAWC, H.E.S.S., MAGIC, and CTA. The steady-state cascade flux produced by UHE gamma rays interacting with the EBL and CMB falls within observational limits for realistic spectral indices in the range \(\gamma = 0.3{-}0.5\) (see Eq.~\ref{bbhSpectr1}). This same range, motivated by gamma-ray constraints, also supports the heavy-tailed neutrino luminosity distribution required to explain rare detections such as KM3-230213A.

Exploring BBH mergers in the vicinity of galactic nuclei across diverse astrophysical environments, we find that although the low energy EM counterpart expected in an AGN disk setting can be bright enough for current X–ray and UV/optical facilities its emission arises from the unresolved nuclear region and typically lacks a distinct point source identity. Absent a unique spectral or temporal fingerprint, such a flare is difficult to distinguish from ordinary AGN variability (e.g., blazar like episodes). By contrast, neutrinos propagate essentially unimpeded, preserve timing and directional information, and thus provide the cleanest messenger of the underlying BBH event when combined with a GW trigger.

We further investigated the gamma-ray signal from the brightest plausible single BBH merger source, assuming the gamma-ray emission shares similar luminosity and spectral characteristics with the associated UHE neutrinos. For events located within the field of view of current gamma-ray facilities, we find that such sources would be detectable up to redshift \(z \approx 0.2\). This implies that more distant brightest BBH neutrino sources could produce events like KM3-230213A without yielding detectable gamma-ray cascades, consistent with the absence of correlated electromagnetic signals.

We explored a possible correspondence between the power-law statistics of percolating Higgs bubble networks and the observed heavy-tailed distribution of neutrino luminosities. In particular, near the percolation threshold, the number of connected bubble clusters exhibits universal scaling behavior, which might be mapped onto the distribution of \(\mu\)BHs and consequently neutrino luminosities produced within the turbulent vacuum region  sandwiched  between the horizons of closely merging BBHs. Within this speculative framework, the emergence of a Pareto-like tail in the emission distribution could reflect features reminiscent of critical behavior in continuum percolation, suggesting a possible unified treatment of BBH merger energetics shaped by vacuum dynamics.

By linking BBH mergers to EW vacuum instability, this work illuminates a potentially rich intersection between gravitational wave astrophysics, high-energy neutrino astronomy, and gamma-ray observations. Looking ahead, coordinated multi-messenger campaigns-particularly those combining gravitational wave, neutrino, and gamma-ray data from next-generation facilities such as IceCube-Gen2 and CTA will be essential for testing the predictions developed here. In particular, the detection of a larger population of faint or transient UHE neutrino sources, along with individual transient gamma-ray counterparts, could validate the heavy-tailed emission statistics implied by the Pareto distribution. Such observations may offer unprecedented insight into EW vacuum instability phenomena at the interface of gravity and particle physics, and potentially provide the first indirect evidence for vacuum decay and BH's Hawking evaporation.

%%%%%%%%%%%%%%%%%%%%%%%%%%%%%%%%%%%%%%%%%%%%%%%%%%%%%%%%%%%%%%%%%%%%%%%%%%%%
%%%%%%%%%%%%%%%%%%%%%%%%%%%%%%%%%%%%%%%%%%%%%%%%%%%%%%%%%%%%%%%%%%%%%%%%%%%%

\medskip

\section*{Acknowledgements}

The work of R.K. was partially supported by the Kakos Endowed Chair in Science Fellowship.

%%%%%%%%%%%%%%%%%%%%%%%%%%%%%%%%%%%%%%%%%%%%%%%%%%%%%%%%%%%%%%%%%%%%%%%%%%%%


\end{document}